\documentclass[aps,prb,twocolumn,showpacs,amsmath,superscriptaddress]{revtex4}
\usepackage{graphics}
\usepackage{epsfig}
\usepackage[colorlinks=true,citecolor=blue,linkcolor=red,linktocpage=true,pagebackref=false]{hyperref}
\usepackage{color,soul} 

\begin{document} 
   \title{ 
   Single-electron source: Adiabatic versus non-adiabatic emission
      }
\author{Michael Moskalets}
\affiliation{D\'epartement de Physique Th\'eorique, Universit\'e de Gen\`eve, CH-1211 Gen\`eve 4, Switzerland}
\affiliation{Department of Metal and Semiconductor Physics, 
NTU ``Kharkiv Polytechnic Institute", 61002 Kharkiv, Ukraine
}

\author{G\'eraldine Haack} 
\affiliation{D\'epartement de Physique Th\'eorique, Universit\'e de Gen\`eve, CH-1211 Gen\`eve 4, Switzerland}
\affiliation{Dahlem Center for Quantum Complex Systems and Fachbereich Physik, Freie Universit\"at Berlin, 14195 Berlin, Germany
}

\author{Markus B\"uttiker}
\affiliation{D\'epartement de Physique Th\'eorique, Universit\'e de Gen\`eve, CH-1211 Gen\`eve 4, Switzerland}

\date\today
 \begin{abstract}
We investigate adiabatic and non-adiabatic emission of single particles into an edge state using an analytically solvable dynamical scattering matrix model of an on-demand source. 
We compare adiabatic and non-adiabatic emissions by considering two geometries: a collider geometry where two emitters are coupled to two different edge states and a series geometry where two emitters are coupled to the same edge state.
Most effects observed for adiabatic emitters also occur for non-adiabatic emitters. In particular this applies to effects arising due to the overlap of wave-packets colliding at a quantum point contact. Specifically we compare the Pauli peak (the fermionic analog of the bosonic Hong-Ou-Mandel dip) for the adiabatic and non-adiabatic collider and find them to be similar. In contrast we find a striking difference between the two operating conditions in the series geometry in which particles are emitted into the same edge state. Whereas the squared average charge current can be nullified for both operating conditions, the heat current can be made to vanish only with adiabatic emitters. 
\end{abstract}
\maketitle

\section{Introduction}
\label{intro}

Recent progress in the field of dynamical quantum transport \cite{Moskalets:2011cw} opens new and fascinating perspectives for exploring and understanding mesoscopic and nanoscopic conductors. 
With the implementation of an on-demand single-electron emitter  \cite{Feve:2007jx,Blumenthal07} not relying on electron-electron interaction it is possible to address directly dynamic properties of a single-electron state in solids. 
The single-particle nature of emitted wave-packets was demonstrated using the noise measurements.\cite{Maire:2008hx,Mahe:2010cp,Bocquillon:2012if,Fricke:2012vk} 
To investigate the coherence properties of emitted wave-packets an approach based  on the measurement of current correlations at a beam splitter \cite{Grenier11,Grenier:2011js} 
and 
an approach based on the measurement of current at the output of an interferometer \cite{HMB12}  
have already been proposed. 

\begin{figure}[t]
\begin{center}
\includegraphics[width=70mm]{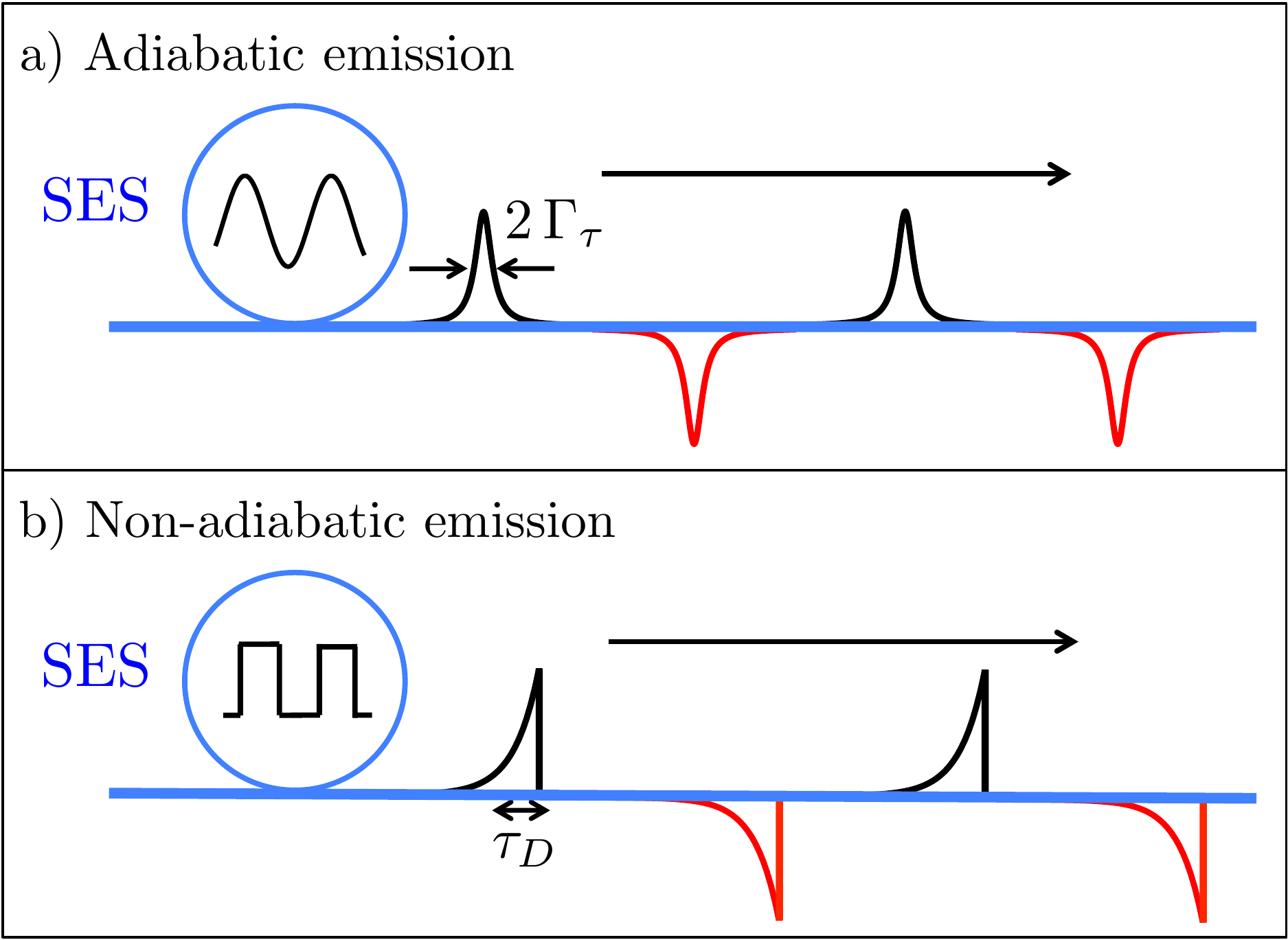}
\caption{
(Color online) Single-electron sources (SES) emit a train of electrons (black pulses) alternating with holes (red pulses) into an edge states (serving as an electronic waveguide). Electron and hole pulses are well separated in time and space. The shape of the single-particle wave-packets depends crucially on the way the source is driven: a) Adiabatic emission: the SES is driven by a smooth periodic potential, the pulse as a function of time has a Lorentzian shape with width $2\Gamma_\tau$. b) Non-adiabatic emission: the SES is driven by a pulsed periodic potential, the pulse as a function of time has an exponential shape characterized by the dwell time $\tau_D$.}
\label{fig1}
\end{center}
\end{figure}

The state of an electron depends crucially on the way it is emitted, see Fig.~\ref{fig1}. 
In most experiments with such high-speed single-electron sources -- see, e.g., Refs.~
\onlinecite{Fujiwara08,Kaestner08,Wright11,Leich11,Fletcher:2012te}, 
also the theoretical proposal in Ref.~\onlinecite{BS11} and the analysis of a single-electron capture in Ref.~\onlinecite{Kashcheyevs:2012km} -- electrons are emitted from the quantum dot with energy  far above the Fermi level. 
On the other hand in theory many effects were predicted for electrons emitted adiabatically almost at the surface of the Fermi sea: The shot-noise quantization \cite{KKL06,OSMB08,KSL08,ZSAM09}; the shot-noise suppression effect \cite{OSMB08,MB11}; a two-particle interference and entanglement generation \cite{SMB09,SSMB10,SASB11} interesting for quantum information applications \cite{Sherkunov:2012dg}; a particle reabsorption \cite{SOMB08,MB09}; the suppression of a single-particle interference by collisions. \cite{JSMoskalets:2011cw} 
Recently also single and few-electron sources based on the generation of Lorentzian voltage pulses applied to a ballistic conductor as proposed in Refs.~\onlinecite{Levitov:1996ie,Ivanov:1997wz}, and discussed in detail in Ref.~\onlinecite{Dubois:2012us}, have now been realized experimentally in Ref.~\onlinecite{Dubois12}. The properties of a single-electron state generated by such a source are similar to those of the state emitted adiabatically by a single lead mesoscopic capacitor.\cite{Feve:2007jx} 

%%%%%%%
%%%%%%%

Our aim here is to answer the question of whether one can expect similar effects with particles emitted non-adiabatically or not. 
For this purpose we analyze the single-electron source of Ref.~\onlinecite{Feve:2007jx} because it can operate in both adiabatic and non-adiabatic emission conditions. 
Moreover it seems that its properties are well described by a non-interacting theory. \cite{Gabelli06,Feve:2007jx,Parmentier11}
That makes it possible to develop a relatively simple analytical theory which describes both operating conditions.

We use a non-interacting model \cite{PTB96,Feve:2007jx,MSB08}, in which the source consists of a single circular edge state, a Fabry-P\'erot cavity, weakly coupled to a linear edge state, which plays the role of an electron waveguide. 
This analytical model is in good agreement with actual experiments.
In the weak coupling limit, the transparency of the quantum point contact connecting the cavity and the electron waveguide is small, $T \ll 1$. 
All relevant energies are smaller than the Fermi energy, $\mu$ and the energy spectrum of electrons can be linearized in the vicinity of $\mu$. 
That results in the equidistant spectrum of the cavity with level spacing $\Delta = h/\tau$ defined by the time of flight, $\tau$, around the circular edge state of the  cavity.
A metallic top-gate with potential $U(t)$ periodically changes the position of the quantum levels in the  cavity. 
We assume optimal operating conditions \cite{Feve:2007jx,Mahe:2010cp,Parmentier11} which require that the Fermi level is positioned in the middle of two levels of the cavity and the gate potential changes with amplitude $\Delta$. 
In this case only one level crosses the Fermi energy: 
When it raises above the Fermi level an electron is emitted from the cavity into the waveguide, whereas when the level sinks below the Fermi level an electron is absorbed by the cavity hence a hole appears in the stream of electrons within the waveguide. Such a source generates no DC current and is often referred to as a quantum capacitor. \cite{BTP93,Gabelli06}

To get an intuitive estimate of both the shape and the duration of a single-particle wave-packet we look at the current pulse emitted by the cavity. 
The sudden change of a potential, $eU(t) = \Delta \theta(t - t_{-})$, results in a transient current pulse (an expectation value),\cite{Feve:2007jx,KSL08,MSB08} 

\begin{eqnarray}
I_{na}(t) = \theta(t - t_{-}) \frac{e}{\tau_{D} }\, e^{-\frac{t - t_{-} }{\tau_{D} }} \,,
\label{pulse-na}
\end{eqnarray}
\ \\
\noindent
with highly asymmetric shape (we ignore a fine structure \cite{MSB08,Sasaoka10} on the scale of $\tau$).
Here $e$ is the electron charge  
and $\theta(t)$ the Heaviside theta-function. The time $t_{-}$ denotes the time at which the potential changes, leading to the emission of an electron and the label ``$na$'' stands for non-adiabatic. Indeed, as explained below, such an emission process corresponds to non-adiabatic emission conditions. The time $\tau_D$,
%, $\theta(t)$ the Heaviside theta-function, and 

\begin{eqnarray}
\tau_{D} = \frac{\tau }{T }  \,,
\label{taud}
\end{eqnarray}
\ \\
\noindent 
the dwell time of an electron in the cavity. Therefore the dwell time 
sets a relevant time-scale of the problem under consideration. 
First, the period ${\cal T}$ of the gate voltage, $U(t) = U(t + {\cal T})$,  should be long enough
for the driven cavity to work as a single-particle source,\cite{Feve:2007jx}

\begin{eqnarray}
 {\cal T} \gg \tau_{D}   \,.
\label{qe}
\end{eqnarray}
\ \\
\noindent  
Note that to operate the source periodically, the energy level needs to be returned back to its initial position by applying the opposite potential $- \Delta \theta(t - t_{+})$. Here, $t_+$ denotes the time at which the emission of a hole starts. The delay between subsequent potential steps should be longer than the duration of a current pulse, $t_{+} - t_{-} \gg \tau_{D}$, to allow an electron emission to be completed: the emitted charge $q = \int_{t_{-}}^{t_{+}} I(t) dt$ should be equal to an electron charge, $q=e$. At time $t_{+}$ a hole can be emitted.  

%Note; to operate periodically we need to return the level back by applying the opposite potential $- \Delta \theta(t - t_{+})$.  The delay between subsequent potential steps should be longer than the duration of a current pulse, $t_{+} - t_{-} \gg \tau_{D}$, to allow an electron emission to be completed: The emitted charge $q = \int_{t_{-}}^{t_{+}} I(t) dt$ should be equal to an electron charge, $q=e$. At time $t_{+}$ a hole can be emitted.  

Second, the dwell time, $\tau_{D}$, defines the condition of adiabatic or non-adiabatic emission.  
%We mark the corresponding quantities with indices ``$ad$'' or ``$na$'', respectively. 
If the potential $U(t)$ changes fast on the scale of $\tau_{D}$, then we speak about a non-adiabatic emission.
In this case the shape of an emitted current pulse is asymmetric and given by Eq.~(\ref{pulse-na}).
In contrast, if $U(t)$ changes smoothly compared to $\tau_D$, the current pulse is predicted to be symmetric. \cite{OSMB08,KSL08}  
%Close to $t_{-}$ it is,
Close to $t_{-}$, the corresponding current pulse $I_{ad}$ reads:

\begin{eqnarray}
I_{ad}(t) =  \frac{e \Gamma_{\tau}/\pi }{\left( t - t_{-} \right)^{2} + \Gamma_{\tau}^{2} } \,.
\label{pulse-a}
\end{eqnarray}
\ \\
\noindent
Now the duration $2\Gamma_{\tau}$ of a current pulse is defined by the  time of crossing, 

\begin{eqnarray}
\Gamma_{\tau} = \frac{\delta  }{ \left|  e\,dU/dt|_{t = t_{-}} \right| } ,
\label{gam}
\end{eqnarray}
\ \\
\noindent
where $2\delta$ is the width of a quantum level in the weakly coupled cavity.  In the model used $\delta = T \Delta /(4\pi)$. 
For $eU(t) = (\Delta/2) \cos(\Omega t)$, where $\Omega = 2\pi/{\cal T}$, and $t_{-} = 3{\cal T}/4$ we find

\begin{eqnarray}
\Gamma_{\tau} = {\cal T} \, \frac{T  }{4 \pi^{2} } \,.
\label{gamma}
\end{eqnarray}
\ \\
\noindent
Remarkably, it was shown in Ref.~\onlinecite{HMSB11} that the pulse duration  $2\Gamma_{\tau}$ also sets the single-particle coherence time of an electron emitted adiabatically. 
This shows that the source, described by this analytical model, has no intrinsic dephasing processes.\cite{Haack_these,HMB12}
This makes the emitted single electron states of particular interest for further applications in quantum information processing.

Equation (\ref{pulse-a}) is calculated assuming that \cite{SOMB08} 

\begin{eqnarray}
\Gamma_{\tau} \gg \tau_{D}  \,.
\label{ad}
\end{eqnarray}
\ \\
\noindent
It means that the level of the cavity crosses the Fermi sea level so slowly that an electron has enough time to leave the cavity once his energy becomes larger than the Fermi energy. 

From Eq.~(\ref{ad}) it also follows that the width of a wave-packet emitted adiabatically is much larger than the width of a wave-packet emitted non-adiabatically. 
Apparently with decreasing crossing time $\Gamma_{\tau}$, keeping the period $\cal T$ large compared to $\tau_{D}$, the shape of the pulse evolves from adiabatic, Eq.~(\ref{pulse-a}), to non-adiabatic, Eq.~(\ref{pulse-na}). For a level driven with a constant speed, an analysis describing this crossover can be found in Ref.~\onlinecite{KSL08}.

The current pulses $I_{na}(t)$, Eq.~(\ref{pulse-na}), and $I_{ad}(t)$, Eq.~(\ref{pulse-a}), have both similar and different features. 
On one hand, they both carry a quantized charge. 
%the similarity is that each of them carries a quantized charge and corresponds to a single particle. 
Therefore, we anticipate that they both should show similar quantization effects \cite{KKL06,OSMB08,KSL08,ZSAM09} and effects arising due to the overlap of wave-packets \cite{OSMB08,MB11,SMB09,SSMB10,SASB11,JSMoskalets:2011cw}. 
We use below the shot noise suppression effect as an example. 

%%%%%%%
\begin{figure}[b]
\begin{center}
\includegraphics[width=70mm]{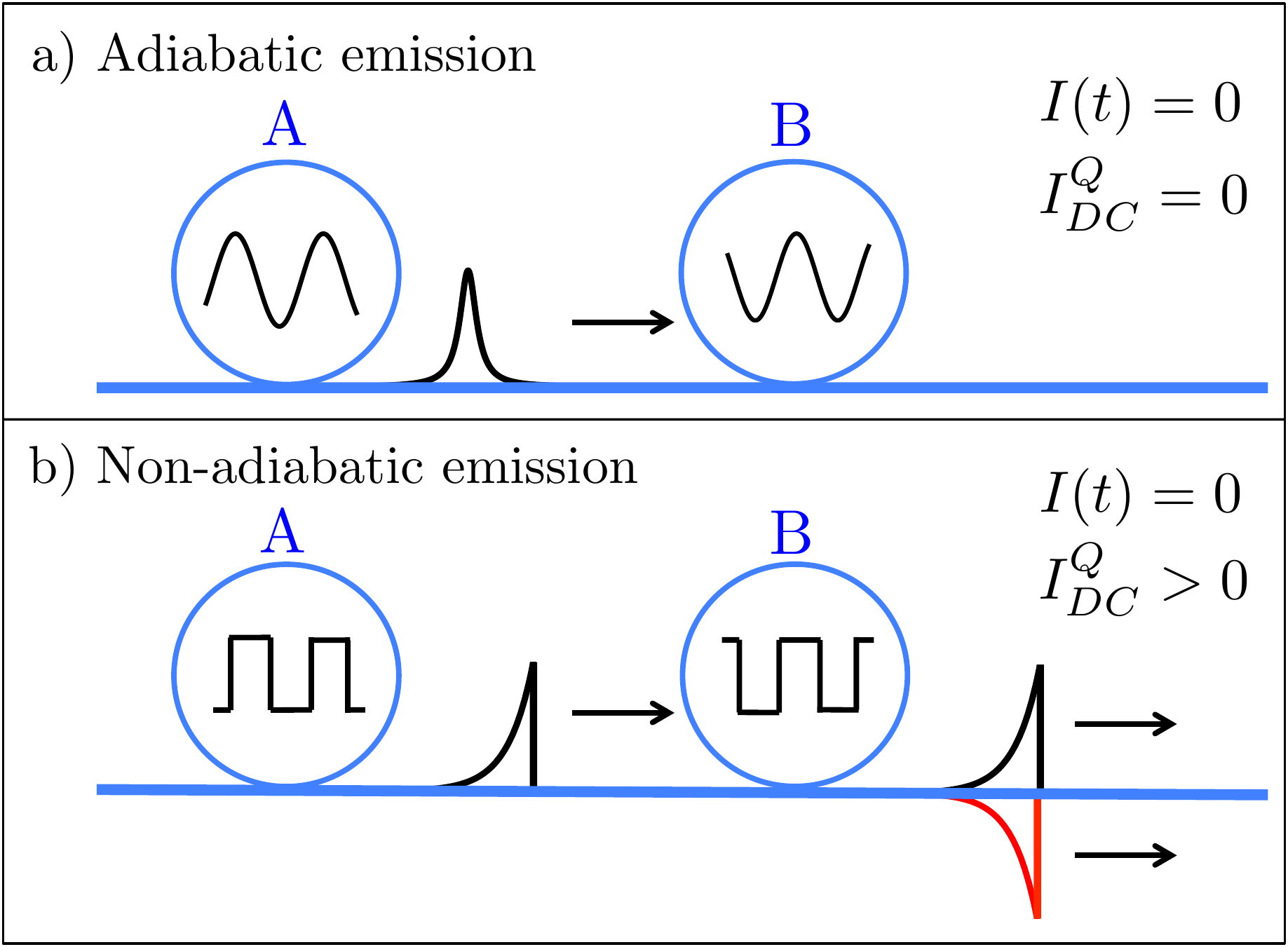}
\caption{
The two-particle emitter consists of two cavities specified by same parameters and coupled to the same chiral edge state. Electrons propagate along edge states shown as blue solid lines. The cavity B is tuned to emit a hole at the time when the electron emitted by cavity A reaches cavity B. (a) Adiabatic emission: when the driven potential is slow and smooth, an electron emitted by the  cavity A is completely reabsorbed by the cavity B. The re-absorption process is the time reversed emission process which is possible because of the symmetric shape of the single-particle states. Both the charge current $I(t)$ and the DC heat flow $I_{DC}^Q$ nullify. (b) Non-adiabatic emission: the cavities are driven by pulsed potentials and electron-hole pairs are emitted. Since this pair is neutral, the time-dependent current is zero, $I(t) = 0$. However, both the electron and the hole carry energy. Because of the asymmetric shape of the pulses reabsorption can not be a time-reversed emission process. There is no absorption effect, the generated DC heat flow is not zero: $I_{DC}^Q > 0$.}
\label{fig2}
\end{center}
\end{figure}
%%%%%%%

On the other hand for some effects the shape of a wave-packet is crucial. 
As an example below we use the effect of reabsorption \cite{SOMB08,MB09} predicted for  the adiabatic regime: If two cavities are coupled to the same edge state, then  the electron emitted adiabatically by one cavity can be reabsorbed by another cavity emitting a hole at the same time, see Fig.~\ref{fig2} (a). 
First of all, in this regime the time-dependent current is zero, $I(t) = 0$. \cite{SOMB08} 
This current consists of two parts, electron, $I^{e}(t)$, and hole, $I^{h}(t)$, which compensates each other: $I^{e}(t) = - I^{h}(t)\, \rightarrow\, I(t) = I^{e}(t) + I^{h}(t) = 0$.
To clarify whether it is merely a compensation effect or a reabsorption effect, additionally the heat generated by the two cavities was analyzed. \cite{MB09} 
It was shown that each particle, either an electron or a hole, carries an excess energy (over the Fermi energy)

\begin{eqnarray}
{\cal E}_{ad} = \frac{\hbar }{2 \Gamma_{\tau} } \,.
\label{en-ad}
\end{eqnarray}
\ \\
\noindent
This energy can be understood as the work done by the potential $U(t)$ on the particle during its escape from the cavity. 
The particle starts to escape when its energy becomes equal to the Fermi energy.
The time it takes to escape is the dwell time, $\tau_{D} = h/(T \Delta)$, given in  Eq.~(\ref{taud}). 
We use Eq.~(\ref{gam}) and find ${\cal E}_{ad} =  \tau_{D} \left|  e dU/dt|_{t = t_{-}}\right| $. 
Notice the energy of a particle in the cavity has an uncertainty $\delta$ (the level width). 
This results in the uncertainty $\Gamma_{\tau}$ of the time when a particle starts to escape the cavity.     
That in turns defines the width of the current pulse, Eq.~(\ref{pulse-a}). 

If two cavities emit an electron and a hole at different times, then these two particles together carry the energy $2{\cal E}_{ad}$. 
However, if an electron and a hole are emitted at the same time (the time of flight between the cavities should be trivially taken into account) then the extra energy flowing out of the system is zero.\cite{MB09} 
Clearly this means that an electron emitted by the cavity $A$ and carrying an energy ${\cal E}_{ad}$  was reabsorbed by the cavity $B$.  
This effect is paradoxical: On one hand, in fact, the hole emission is an electron absorption. On the other hand, the cavity $B$ can absorb any  electrons in the waveguide passing it. Why does it absorb the electron emitted by the  cavity $A$?

Possibly this effect can be understood using time-reversal symmetry arguments. 
First, let us take only one cavity and let it emit an electron. 
After that let us reverse time. 
Apparently the emitted electron will be reabsorbed. 
Importantly, the portion of the wave-packet emitted last will be reabsorbed first. 
Now  let us take two identical cavities and let us drive them with potentials $U_{1}(t)$ and $U_{2}(t)$ related by the time-reversal symmetry, $U_{2}(t) = U_{1}(-t)$. 
Note with such potentials if the first cavity emits an electron the second cavity emits a hole and vice versa. We can expect the second cavity to be an analogue of the time-reversal twin of the first cavity. 
To make such an analogy complete, the shape of the wave-packet does matter. 
Because the second cavity will first reabsorb (if possible) the part of the wave-packet, which was emitted first. 
In contrast, the true time-reversal twin will first absorb what was emitter last. 
If the shape of a wave-packet is symmetric, as in the adiabatic emission regime, (for the corresponding current pulse see Eq.~(\ref{pulse-a})), then there is no difference between what was emitted first and what was emitted last.
Consequently the second cavity can play the role of the time-reversal twin and reabsorb what was emitted by the first cavity. 

However, if the shape of a wave-packet is non-symmetric, as in the non-adiabatic emission regime, (for the corresponding current pulse see Eq.~(\ref{pulse-na})), then there is a striking difference between what was emitted first and last. 
As a consequence what the second cavity sees is different from what the time-reversal twin would see. 
Therefore, adiabatic and non-adiabatic cavities work differently. 
As we show below, in the non-adiabatic emission regime both cavities emit together an electron-hole pair, which carries no charge, $I(t) = 0$, but carries a non-zero energy.   

The paper is organized as follows: 
In Sec.~\ref{shot} we discuss the shot noise quantization and the shot noise suppression effect for electrons emitted non-adiabatically. 
In Sec.~\ref{two} the DC heat flow generated by the two-particle emitter working in the non-adiabatic regime is analyzed. 
We conclude with a brief discussion in Sec.~\ref{concl}.
Details of the calculations are in appendices. 
In Appendix \ref{scat} we derive the Floquet scattering amplitude for a cavity driven by the pulsed potential. 
In Appendix \ref{shn} the zero frequency correlation function for currents flowing through the electron collider circuit is found. 
In Appendix \ref{twope} we discuss the DC heat flow generated by the two-cavity emitter.

\section{Pauli suppression of shot noise}
\label{shot}

\begin{figure}[b]
\begin{center}
\includegraphics[width=80mm]{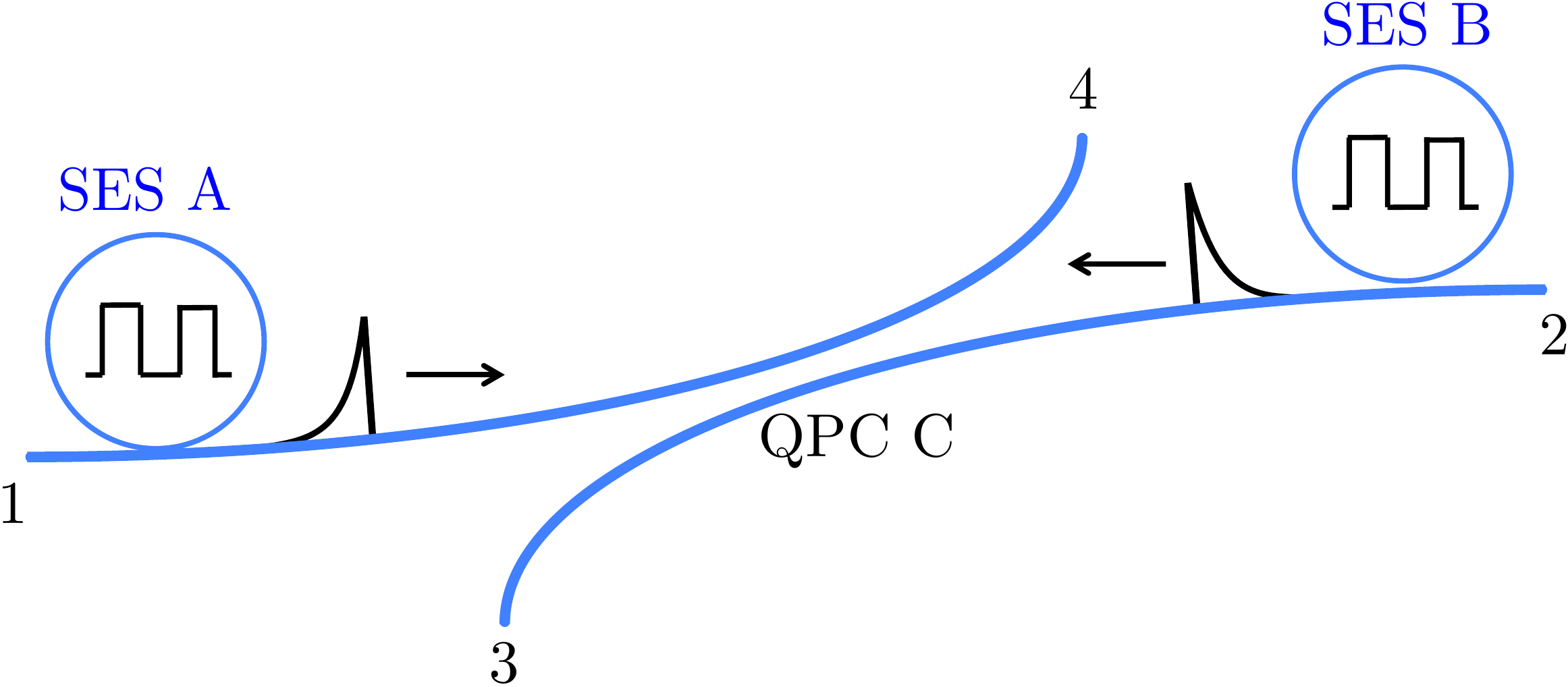}
\caption{
(Color online) A mesoscopic electron collider. Two single-electron sources, $A$ and $B$, driven by pulsed potentials emit electrons into the waveguides. 
The emitted electrons collide at the quantum point contact C. 
The metallic reservoirs are labeled by the number from $1$ to $4$.
}
\label{fig3}
\end{center}
\end{figure}

A mesoscopic electron collider is a circuit in which electrons incident from different leads can meet and collide \cite{Buttiker:1992vr,Liu:1998wr,Blanter:2000wi,Feve:2008im} at a wave-splitter (a mesoscopic quantum point contact).  
We consider a collider with two single-electron emitters, $A$ and $B$, weakly coupled to the  two chiral edge states, the electron waveguides, which are in turn coupled via a quantum point contact C (QPC C) with transmission probability $T_{C}$, Fig.~\ref{fig3}. 
For simplicity we use the cavities with identical parameters but emitting particles possibly at different times.   
Four metallic reservoirs, $1$ to $4$, are kept at the same electrochemical potential $\mu$ and zero temperature.

%%%%%%%
%%%%%%%

Each source, $j = A\,, B$, is driven by the periodic, $U^{j}(t) = U^{j}(t + {\cal T})$, potential with amplitude $\Delta/e > 0$.
The minimal value $U_{0}$ is chosen such that the Fermi level lies exactly in the middle between the two quantum levels of the cavity:

\begin{equation}
U^{j}(t) = \left\{
\begin{array}{ll}
U_{0} \,,  &-{\cal T}/2 < t <  t^{j}_{-}\,,\\
 &  \\
U_{0} + \Delta/e \,,  &t^{j}_{-} < t <  t^{j}_{+}\,,\\
 &  \\
U_{0} \,,  & t^{j}_{+} < t <  {\cal T}/2\,.\\
 &  \\
\end{array}
\right.
\label{pulse0} 
\end{equation}
\ \\
\noindent
We are interested in the zero-frequency correlation function \cite{Blanter:2000wi}, ${\cal P}_{34}$,  for currents outgoing to contacts $3$ and $4$. 
These currents are due to the particles, electrons and holes, emitted by the sources at times $t^{j}_{-} + a{\cal T}$ and $t^{j}_{+} + a{\cal T}$, respectively. Here $a$ is an integer.

To calculate ${\cal P}_{34}$ we use the theory of Ref.~\onlinecite{MBnoise04} with the central object being the Floquet scattering matrix of the source $j$, $\hat S_{F}^{j}$. 
In our model this is a matrix in the energy space with elements $S_{F}^{j}(E_{n},E)$, where $E_{n} = E + n\hbar\Omega$, which are photon-assisted  amplitudes for an electron in the waveguide to pass through the cavity $j$ and to absorb (for $n>0$) or emit (for $n<0$) $n$ energy quanta $\hbar\Omega$. 

It is convenient to introduce the scattering amplitude $S_{in}^{j}(t,E)$ whose Fourier coefficients are $S_{in,n}^{j}(E) = S_{F}^{j}(E_{n},E)$, see Eq.~(\ref{fla}).
For the model of the source introduced above we have (see Eq.~(\ref{inn}) and Appendix~\ref{scat} for details of calculations): \cite{HMB12}

\begin{eqnarray}
S_{in,n}^{j}(E) &=&  S(E)  \,\frac{\sin\left( \pi \frac{n \hbar\Omega }{\Delta } \right) }{\pi n   } \, e^{ i \pi \frac{n \hbar\Omega  }{\Delta }  } \nonumber \\
\label{sin} \\
&& \times \left\{  \delta_{n,0} - \frac{  \frac{ e^{i n \Omega t^{j}_{-}} }{1 -  \frac{n\hbar\Omega }{\Delta }  } + \frac{ e^{i n \Omega t^{j}_{+}} }{1 +  \frac{n\hbar\Omega }{\Delta } } }{\rho^{*}(E) \rho(E_{n}) }  \right\}  . \nonumber 
\end{eqnarray}
\ \\
\noindent
Here $S(E)$ is the stationary scattering amplitude of the cavity with potential $U_{0}$, see Eq.~(\ref{froz}) with $U(t) = U_{0}$, and $\rho(E) = \left[ 1 - \sqrt{1-T} e^{ i\phi(E) } \right]/\sqrt{T} $ with $\phi(E) = \pi + 2\pi (E-\mu)/\Delta$ a phase picked up by an electron with energy $E$ during one turn around the cavity.

\subsection{Quantized noise of a single source}

Let us for a moment switch off one of the sources.
Then we find, see Appendix \ref{shn}, Eq.~(\ref{noise-1-2}), 

\begin{eqnarray}
{\cal P}_{34}^{na} = - 2 {\cal P}_{0} \,,
\label{sh-1}
\end{eqnarray}
\ \\
\noindent
where ${\cal P}_{0} = e^{2} T_{C} (1-T_{C})/{\cal T}$. 
This result coincides completely with the one found for the cavity emitting particles adiabatically \cite{OSMB08} and, therefore, it tells us that at zero temperature the zero frequency current correlation function is independent of the parameters of both the cavity and driving potential as far as the cavity emits separate particles. 
The quantity ${\cal P}_{34}^{na}$, Eq.~(\ref{sh-1}), at zero temperature can be interpreted as due to the shot noise of two indivisible quanta, one electron and one hole, emitted during each period and scattered at the quantum point contact C to either the contact $3$ or $4$. 
Such a partition noise was measured in Ref.~\onlinecite{Bocquillon:2012if}. 
The deviation from the theoretical prediction found is attributed to the effect of a non-zero temperature. 

If the source emits $N$ electron and $N$ holes during the period then the factor $2$ in Eq.~(\ref{sh-1}) is replaced by the factor $2N$. 
We also note that the noise per particle, $-{\cal P}_{0}$, is just the result of the partition noise of a dc-source biased with the voltage $eV = \hbar\Omega$, see e.g. Ref.~\onlinecite{Blanter:2000wi}.

Let us now consider the situation where the sources $A$ and $B$ are both operating as shown in Fig. \ref{fig3}.

\subsection{Shot noise suppression effect}

\begin{figure}[t]
\begin{center}
\includegraphics[width=80mm]{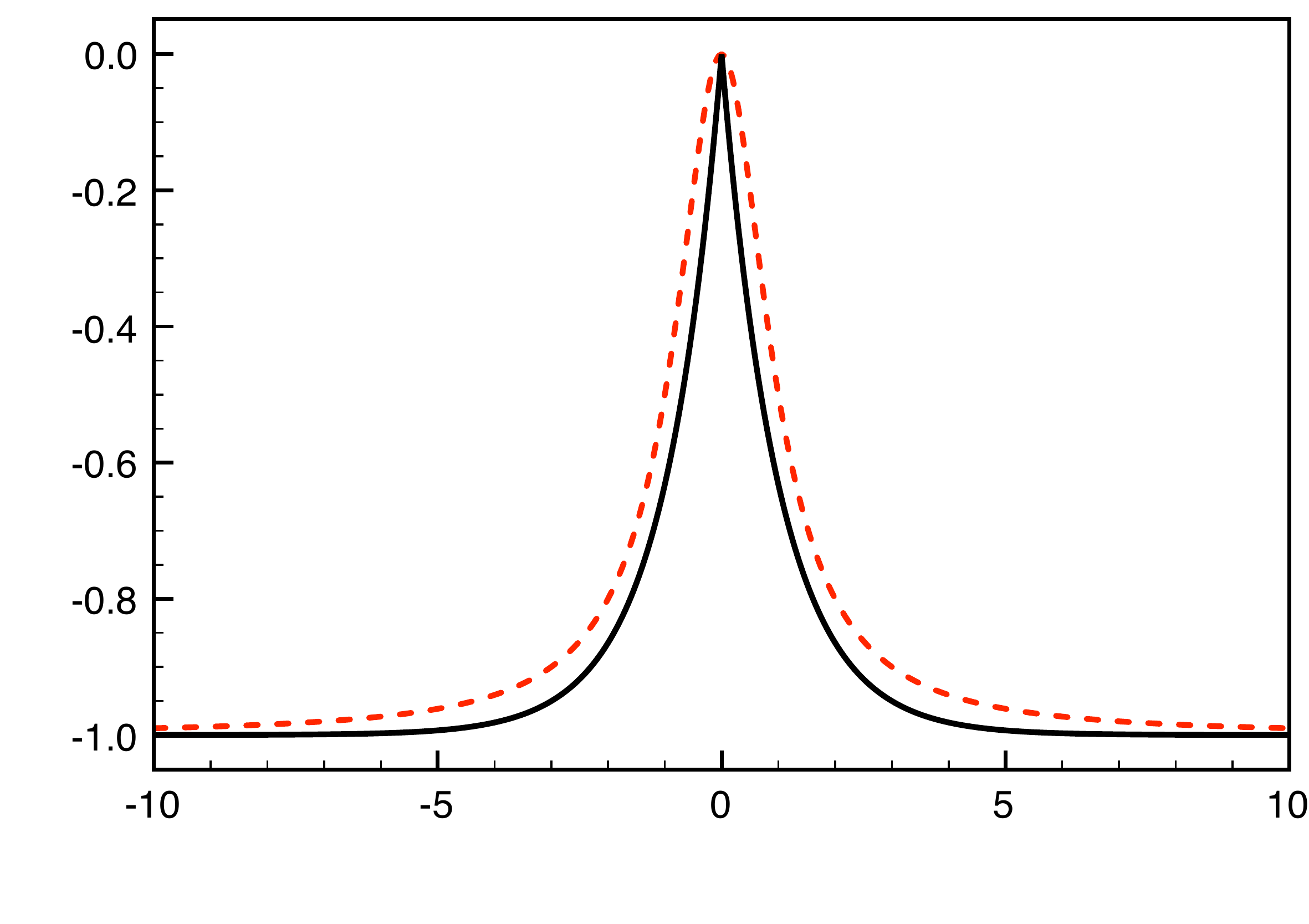}
\begin{picture}(0,0)
\put(-230,135){${\cal P}$}
\put(-30,5){$\delta t$}
\end{picture}
\caption{
(Color online) The Pauli peak: The shot noise per particle of an electron collider, ${\cal P} = {\cal P}_{34}/(2N {\cal P}_{0})$, as a function of the time delay $\delta t \equiv t_{-}^{A} - t_{-}^{B} =  t_{+}^{A} - t_{+}^{B}$ normalized by $\Gamma_{\tau}$ for an adiabatic emission, Eq.~(\ref{a-sh-2}), (red dashed line) and by $\tau_{D}$  for a non-adiabatic emission, Eq.~(\ref{sh-2}), (black solid line). 
}
\label{fig4}
\end{center}
\end{figure}
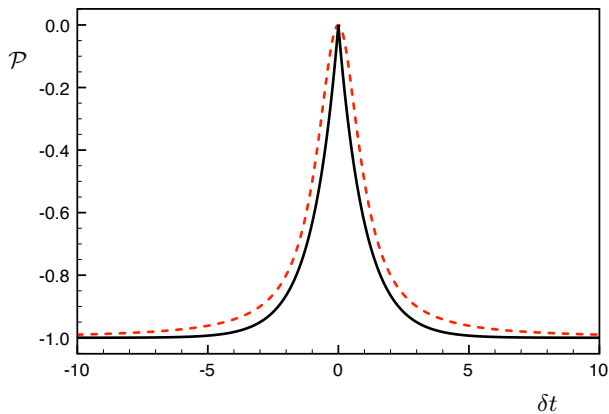

If both sources work then the total shot noise depends crucially on whether two electrons (respectively two holes) emitted by the different sources pass the QPC C  at different times or not. 
If the particles pass the quantum point contact C at different times, $|t_{\mp}^{A} - t_{\mp}^{B}| \gg \tau_{D}$, then the shot noise is, ${\cal P}_{34}^{na} = - 4 {\cal P}_{0}$, since both sources together emit $4$ particles, two electrons and two holes, during each period. 
Due to the Pauli principle the noise is reduced when particles arrive nearly simultaneously at the QPC. 
This leads to the Pauli peak for the cross-correlator, see  Fig.~(\ref{fig4}), or the Pauli dip in the auto-correlator. 
The Pauli peak is the fermionic analog of the bosonic Hong-Ou-Mandel\cite{Hong:1987gm} dip known in optics.  
We describe the aforementioned reduction with a function $D(\delta t)$ dependent on the difference of arrival times $\delta t = t_{\mp}^{A} - t_{\mp}^{B}$.  
The calculations presented in Appendix \ref{shn-2} give for the non-adiabatic case: 
%(the superscript ``$na$''):

\begin{subequations}
\label{sh-2}
\begin{eqnarray}
\frac{ {\cal P}_{34}^{na}  }{ {\cal P}_{0} } &=&  - 2  D^{na}\left( t_{-}^{A} - t_{-}^{B} \right)  - 2  D^{na}\left( t_{+}^{A} - t_{+}^{B} \right)   , 
\label{sh-2A}
\end{eqnarray}
\begin{eqnarray}
D^{na}\left( \delta t \right) &=& 1 - e^{- \frac{\left| \delta t \right |}{\tau_{D} } }, 
\label{sh-2B}
\end{eqnarray}
\end{subequations}
\ \\
\noindent
where we chose the sources to be placed the same distance from the QPC C. 
Remember we assumed that the two cavities emit wave-packets of the same shape. 
For the case of cavities emitting non-adiabatically wave-packets of different shape see Ref.~\onlinecite{Jonckheere:2012cu}.

The behavior of the shot noise discussed above qualitatively agrees with what we predicted for emitters working under adiabatic conditions: \cite{OSMB08} 
%(the superscript ``$ad$''): \cite{OSMB08}

\begin{subequations}
\label{a-sh-2}
\begin{eqnarray}
\frac{ {\cal P}_{34}^{ad}  }{ {\cal P}_{0} } &=&  - 2  D^{ad}\left( t_{-}^{A} - t_{-}^{B} \right)  - 2  D^{ad}\left( t_{+}^{A} - t_{+}^{B} \right)   , 
\label{a-sh-2A}
\end{eqnarray}
\begin{eqnarray}
D^{ad}\left(\delta t \right) &=& 1 -  \frac{4  \Gamma_{\tau}^{2} }{\left( \delta t \right)^{2} + 4\Gamma_{\tau}^{2} } . 
\label{a-sh-2B}
\end{eqnarray}
\end{subequations}

The reduction function $D\left(\delta t \right)$ can also be calculated from the overlap of wave functions of colliding particles at the quantum point contact (the wave splitter).\cite{Loudon,Blanter:2000wi} 
If the two sources emit wave packets of the same shape then the overlap can be {\it formally}  expressed in terms of the single-particle correlation function $G^{(1)}_{e}\left( t_{1},t_{2} \right)$ discussed in Ref.~\onlinecite{HMB12}.
The reduction function then can be written as follows~:

\begin{eqnarray}
D\left( \delta t \right) = 1 -  v_{D}^{2} \left|  \int dt G^{(1)}_{e}\left( t + \delta t , t  \right) \right|^{2} \,,
\label{dump}
\end{eqnarray}
\ \\
\noindent  
where the integral runs over the time interval when the particles pass the QPC and $v_{D}$ is the velocity of an electron evaluated at the Fermi energy $\mu$. 
The factor $v_{D}^{2}$ is introduced to account for the wave-function normalization in such a way that at the complete overlap, $\delta t = 0$, the reduction function $D=0$.  
For the single-electron source of Ref.~\onlinecite{Feve:2007jx} we found \cite{HMB12} for the adiabatic emission~:

\begin{equation}
G^{(1)}_{e,ad} (t + \delta t , t  ) = \frac{1}{\pi \Gamma_{\tau} v_{D} } \frac{1}{\left(1 - i\frac{ t + \delta t }{\Gamma_{\tau} } \right)\left(1 + i\frac{ t  }{\Gamma_{\tau} } \right) } \,,
\label{eq:coherence}
\end{equation}
\ \\
\noindent
and for the non-adiabatic emission

\begin{eqnarray}
G^{(1)}_{e,ad}(t + \delta t , t) =e^{-i\frac{\Delta }{2 } \frac{\delta t}{\hbar } }  \theta(t) \theta(\delta t ) \frac{ e^{ -\frac{t + \delta t/2 }{\tau_{D} }  } }{ \tau_{D} v_{D}}   \,.    
\label{cohna} 
\end{eqnarray}
\ \\
\noindent
By inserting Eq.~(\ref{eq:coherence}) and Eq.~(\ref{cohna}) in Eq.~(\ref{dump}), we recover respectively the reduction factor in the adiabatic regime, Eq.~(\ref{a-sh-2B}), and in the non-adiabatic regime, Eq.~(\ref{sh-2B}).

In Fig.~\ref{fig4} we show ${\cal P}_{34}^{na}$ and ${\cal P}_{34}^{ad}$ as function of the time difference $\delta t \equiv \delta t_{-} = \delta t_{+}$ (with $ \delta t_{\chi} = t_{\chi}^{A} - t_{\chi}^{B}$, $\chi = \mp$)  normalized on $\tau_{D}$ and $\Gamma_{\tau}$, respectively. 
The Pauli peaks are remarkably similar under the two limiting operating conditions despite the fact that the emitted states are very different. 
However the two operating conditions can be perhaps differentiated experimentally taking a closer look at the top of a peak. It is sharp in the non-adiabatic case and smoother in the adiabatic case. 
We remark that for different incident states the Pauli peak has an asymmetric shape in the non-adiabatic case\cite{Jonckheere:2012cu} but remains symmetric in the adiabatic case\cite{OSMB08}.
Measurements on an electronic collider have now succeeded in detecting the Pauli peak.\cite{Bocquillon:2013dp}

%%%%%%%
%%%%%%%

\section{Two-particle emitter}
\label{two}

In this section we consider the circuit with two cavities connected in series to the {\it same} edge state as shown in Fig.~\ref{fig2} (b). 
If both cavities emit particles at close times, such a circuit serves as a two-particle emitter. 
Its work under the adiabatic condition of emission was analyzed in Refs.~\onlinecite{SOMB08,MB09}. 
Here we analyze it under the non-adiabatic condition of emission when each cavity is driven by the pulsed potential, see Eq.~(\ref{pulse0}). 

We calculate the DC heat flow, $I_{DC}^{Q}$, generated by the two-cavity source as a quantity able to differentiate various cases: (i) Separate emission of particles, (ii) electron-hole emission, (iii) two-electron (two-hole) emission. 
The details of calculations are given in Appendix \ref{twope}. 
Here we discuss the results. 

We start from conditions when the cavities emit particles at different times, see Appendix \ref{heatind}.  
The DC heat flow, $I_{DC}^{Q} = 4 {\cal E}_{na}/{\cal T}$, is due to four particles emitted by both cavities during each period. Each particle carries  an excess (over the Fermi sea level) energy

\begin{eqnarray}
{\cal E}_{na} = \frac{\Delta }{2 }.
\label{en-na}
\end{eqnarray}
\ \\
\noindent 
The above result is clear, since the potential $U^{j}(t)$, Eq.~(\ref{pulse0}), moves a quantum level of the cavity $j$ by $\Delta/2$ above (below) the Fermi energy when an electron (a hole) is emitted. 

The single-particle energy ${\cal E}$ can be also understood on the base of the Joule-Lenz law,

\begin{eqnarray}
{\cal E} = R \int dt  I^{2}(t)  \,.
\label{JL}
\end{eqnarray}
\ \\
\noindent
Here we integrate over a single-particle current pulse.
Under the adiabatic emission condition we use Eq.~(\ref{pulse-a}) for $I(t)$ and Eq.~(\ref{en-ad}) for ${\cal E}$ and find from Eq.~(\ref{JL}) that the relevant resistance,  $R_{ad} \equiv R_{q} = h/(2e^{2})$, is the charge relaxation resistance quantum \cite{BTP93,Nigg:2006kl,Petitjean:2009je,Mora:2010hw,Kashuba:2012fs} which appears in the linear response (admittance)  of the cavity \cite{Gabelli06} (a quantum capacitor) at  low-temperature. 
Under the non-adiabatic emission condition we use Eq.~(\ref{pulse-na}) for $I(t)$ and Eq.~(\ref{en-na}) for ${\cal E}$.
From Eq.~(\ref{JL}) we then find $R_{na} = h/(e^{2} T)$. 
This is an ordinary two-terminal resistance of the (single channel) quantum point contact connecting the cavity to the electron waveguide.  
This resistance was found experimentally in the optimal operation conditions \cite{Feve:2007jx} and it appears in theory in both the high-temperature \cite{MSB08} and incoherent \cite{Nigg08} case. 
Therefore, we see that the factor $R$ in Eq.~(\ref{JL}) is not universal but it depends crucially on the way how an electron is emitted.

Note also that according to Ref.~\onlinecite{Avron:2001fl} the adiabatic source is optimal in the sense that it dissipates the minimal heat per generated particle (an electron or a hole), $R_{ad} = R_{q}$. 
In contrast the non-adiabatic source dissipates more energy, $R_{na} \gg R_{q}$ and thus it is non optimal. 
It is worthwhile to mention that the criteria for an optimal pump generating a DC quantized current\cite{Avron:2001fl} works also in our case for the emitter which produces a quantized AC current, a sequence of alternating electrons and holes.    

Now we come back to the two-particle emitter.  
If two cavities emit two electrons (two holes) simultaneously, see Appendix \ref{heatee}, the energy carried by the pair of particles, ${\cal E}_{na}^{ee} = {\cal E}_{na}^{hh}$, is enhanced two times compared to the condition of independent emission (when two separate particles  carry energy $2{\cal E}_{na}$):

\begin{eqnarray}
{\cal E}_{na}^{ee} = 4 {\cal E}_{na} \,.
\label{ee}
\end{eqnarray}
\ \\
\noindent
The same two-time enhancement was found under the adiabatic emission condition. \cite{MB09}
The enhancement factor two can be understood using the Joule-Lenz law, Eq.~(\ref{JL}), since if the two particles are emitted simultaneously, then  the amplitude of a current pulse is doubled. 
The reason that an energy enhancement can not be avoided follows from the Pauli blocking: The cavity B can not emit a particle with energy ${\cal E}$ if there is a particle with the same energy (emitted by the cavity A). 
Therefore, the cavity B has to emit a particle with an enhanced energy. 
Under the non-adiabatic condition the cavity B excites an electron to the next available quantum level of the cavity and then an electron having energy $\Delta/2 + \Delta$ leaves a cavity. 
This scenario agrees with a non-adiabatic excitation of an electron in a dynamical quantum dot observed in Ref.~\onlinecite{Kataoka11}. 
The direct spectroscopy of energies of electrons emitted by the two-particle source can be done in the same way as proposed in Ref.~\onlinecite{Battista:2012db} for the single-particle emitter. 

The last operating condition we want to discuss is, see Appendix \ref{heateh},  when one cavity emits an electron at the time the other one emits a hole, see Fig.~\ref{fig2}  (b). 
We find that the DC heat flow is not changed compared to the case when the particles are emitted at different times. 
This means that now our source emits electron-hole pairs each carrying finite heat ${\cal E}^{eh}_{na} = 2{\cal E}_{na}$ but zero charge. 
This is in striking contrast with the adiabatic emission case when a  particle emitted by one source is reabsorbed by the other source thus nullifying both the charge current and the DC heat flow. \cite{SOMB08,MB09} 
The nullification of the DC heat flow can also be understood as a work transfer between the external forces \cite{Arrachea:2007ih,Arrachea:2012vb} driving the two particle sources. 
We remark that  in the electron-hole emission case the Joule-Lenz law, Eq.~(\ref{JL}), holds under the adiabatic condition, whereas it seems to be violated under the non-adiabatic condition. 
Note that also the fluctuation-dissipation relation is broken in  the adiabatic operating conditions when the two cavities generate electron-hole pairs. \cite{MB09}

\section{Conclusion}
\label{concl}

Here we developed an analytical Floquet scattering matrix approach to describe the chiral single-electron source driven by the pulsed potential and, therefore, emitting particles, electrons and holes, non-adiabatically. 
We analyzed an electronic collider and the two-particle emitter circuits with such sources and compared them to the analogous circuits with single-electron emitters driven by the smooth potential, i.e., emitting particles adiabatically.  

We found that the collision of electrons approaching a quantum point contact from different sides, see Fig.~\ref{fig3}, suppresses the shot noise. 
This effect is similar to the one found under the adiabatic emission condition \cite{OSMB08} and it is due to the Pauli repulsion between the overlapping electrons, which forces them to go to the different out-puts thus regularizing the outgoing particle flows. 
The sharper suppressing factor, Fig.~\ref{fig4} (black solid line), is due to the spatial asymmetry of traveling wave-packets generated non-adiabatically. 

A more striking difference was found for a circuit comprising two cavities  connected to the same edge state, Fig.~\ref{fig2}. 
If both cavities emit particles at close times such a circuit serves as a two-particle emitter. 
The difference between adiabatic and non-adiabatic emission conditions appears when cavities emit particles of different kind, i.e., one cavity emits a hole at the same time as the other cavity emits an electron. 
If particles are emitted adiabatically, then the cavity B reabsorbs \cite{SOMB08,MB09} what was emitted by the cavity A, whereas in the non-adiabatic emission mode a neutral electron-hole pair having a finite energy is emitted. 
This can be verified by looking at the DC heat flowing out of the system: It is zero under the adiabatic emission condition but finite under the non-adiabatic one. 
If both cavities emit particles of the same kind (two electrons or two holes) then under either adiabatic or non-adiabatic emission conditions we found doubling of heat compared to the case when all particles are emitted at different times. 

Surprisingly the Joule-Lenz law relating a current through and heat released in the macroscopic conductor also holds for the single-particle excitation: The square of the single electron (hole) current pulse  integrated over time gives the heat carried by this particle from the source and released in the macroscopic reservoir. 
This law works under either adiabatic or non-adiabatic emission conditions though with different relevant resistances.
However it is violated completely for the two-particle source emitting an electron-hole pair under the non-adiabatic emission condition.

\begin{acknowledgments}
G. H. acknowledges the support of the Alexander von Humboldt Foundation in the framework of the Alexander von Humboldt Professorship, endowed by the Federal Ministry of Education and Research. 
This work was supported by the Swiss National Science Foundation and Swiss National Center for Competence in Research on Quantum Science and Technology QSIT. 
\end{acknowledgments}

\appendix

\section{The Floquet scattering matrix}
\label{scat}

In Ref.~\onlinecite{MSB08} the scattering amplitude $S_{in}(t,E)$ for an electron with a linear dispersion being scattered off a one-dimensional circular edge state (a cavity) was calculated.
The cavity is driven by the uniform in space and periodic in time potential, $U(t) = U(t + {\cal T})$. 
This amplitude can be presented as the sum of partial amplitudes classified by the number of turns $q$ made by an electron with energy $E$ entering the cavity before leaving it at time $t$:

\begin{equation}
S_{in}(t,E) = r + \tilde{t}^2 \sum_{q=1}^{\infty}r^{q-1} 
e^{i\left\{q\varphi(E) - \Phi_{in,q}(t)\right\}} .
\label{in}
\end{equation}
\ \\
\noindent
Here $r$/$\tilde t$ is the  reflection/transmission amplitude of the quantum point contact connecting the cavity and the chiral one-dimensional conductor (an electron waveguide), $\varphi(E) = \varphi(\mu) + (\tau/\hbar) (E - \mu)$ is the kinematic phase acquired by an electron during one turn in the cavity, $\Phi_{q}$ is the phase due to the time-dependent potential acquired by an electron during $q$ turns,

\begin{eqnarray}
\Phi_{in,q}(t)=\frac{e}{\hbar}\int\limits_{t-q\tau}^{t}dt^{\prime}U(t^{\prime})\,,
\label{phi-in}
\end{eqnarray}
\ \\
\noindent
where $\tau$ is a duration of one turn. 
Details of the derivation can be found in Ref.~\onlinecite{Moskalets:2011cw}

The Fourier coefficients of $S_{in}$ define the elements of the Floquet scattering matrix (in the energy space),

\begin{subequations}
\label{fl}
\begin{eqnarray}
S_{F}(E + n\hbar\Omega, E) = S_{in,n}(E) \equiv \int\limits_{0}^{\cal T} \frac{dt}{\cal T} e^{in\Omega t} S_{in}(t,E) , \nonumber \\
\label{fla}
\end{eqnarray}
\ \\
\noindent
which are amplitudes corresponding to  photon-assisted scattering with exchange of $n$ energy quanta $\hbar\Omega$ between an electron and the driving field.  
Here $\Omega = 2\pi/{\cal T}$ is the frequency of the drive. 
For $n>0$ the electron absorbs energy whereas for $n<0$ it emits energy. 

For some calculations the dual amplitude $S_{out}(E,t)$, which fixes the time when an electron enters the dot \cite{MB08}, is more natural to use. 
Its Fourier coefficients relate to the Floquet scattering amplitudes in the following way:

\begin{eqnarray}
S_{F}(E, E - n\hbar\Omega) \!=\! S_{out,n}(E) \equiv\!\!  \int\limits_{0}^{\cal T} \frac{dt}{\cal T} e^{in\Omega t} S_{out}(E,t) . \nonumber \\
\label{flb}
\end{eqnarray}
\end{subequations}

The amplitudes $S_{in}(t,E)$ and $S_{out}(E,t)$ are generally interrelated in a simple manner. \cite{MB08} 
In particular, for the model of interest here, the amplitude $S_{out}(E,t)$ is given by Eq.~(\ref{in}) with $\Phi_{in,q}(t)$ replaced by 

\begin{eqnarray}
\Phi_{out,q}(t)=\frac{e}{\hbar}\int\limits_{t}^{t+q\tau}dt^{\prime}U(t^{\prime})\,.
\label{phi-out}
\end{eqnarray}

Depending on the ratio between the time of a single turn $\tau$ and a characteristic time during which the driving potential $U(t)$ changes we distinguish adiabatic and non-adiabatic operating conditions.

\subsection{Adiabatic emission}

If the potential $U(t)$ changes slowly, i.e., the maximum relevant frequency is much smaller than $\tau^{-1}$. Thus  we can keep $U(t^{\prime})$ constant while integrating over time in Eq.~(\ref{phi-in}) or Eq.~(\ref{phi-out}). 
We arrive at the {\it frozen} \cite{Avron:2001fl, Moskalets:2011cw} scattering amplitude, $S(U[t];E) \equiv S_{in}(t,E) = S_{out}(E,t)$: 

\begin{subequations}
\label{froz}
\begin{eqnarray}
\!\!\!\!S(U[t];E) = - \,e^{ i(\phi(U[t];E) +\theta_{r} ) } \frac{ 1 - \sqrt{R} e^{-i\phi(U[t];E)} }{ 1 - \sqrt{R} e^{i \phi(U[t];E)} } \,,
\label{froza}
\end{eqnarray}
\ \\
\noindent
where 

\begin{eqnarray}
\phi(U[t];E) = \theta_{r} + \varphi(\mu) + 2\pi \frac{ E - \mu }{ h/\tau}  - 2\pi \frac{ eU(t) }{ h/\tau}  \,,
\label{frozb}
\end{eqnarray}
\end{subequations}
\ \\
\noindent
$\sqrt{R}$ and $\theta_{r}$ are the absolute value and the phase of the reflection amplitude, $r = \sqrt{R} \exp(i \theta_{r})$.

%%%%%%%
\begin{figure}[b]
\begin{center}
\includegraphics[width=85mm]{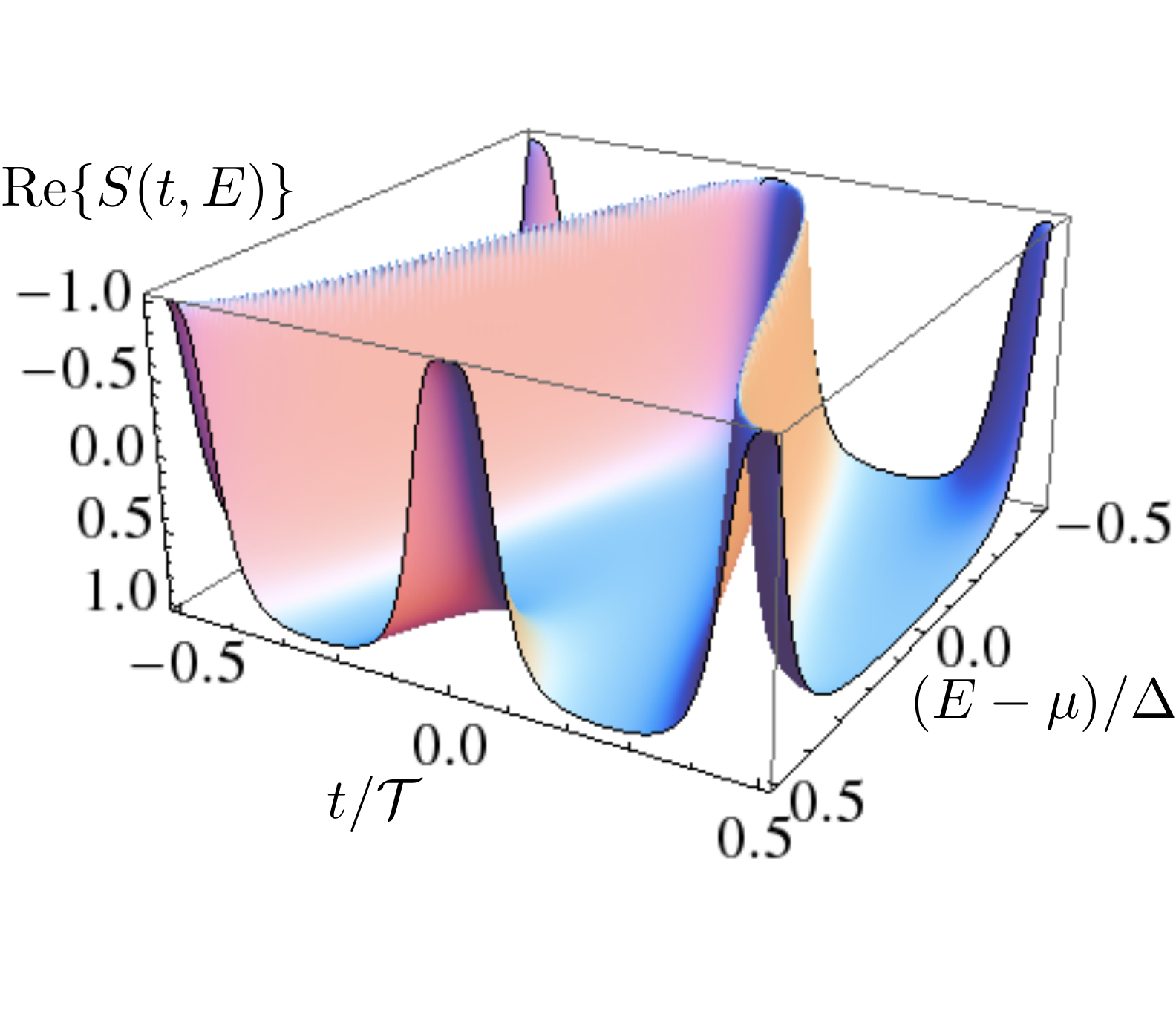}
\caption{
(Color online) Adiabatic emission: The real part of $S\left( t,E \right)$, Eq.~(\ref{froza}), is shown. The time $t$ is measured in units of the period of the drive ${\cal T} = 2\pi/\Omega$. The energy $E$ is measured from the Fermi energy $\mu$ in units of the level spacing $\Delta$. Only one period for both $t$ and $E$ is shown. The transmission probability of a quantum point contact connecting the cavity is $T=0.5$. Other parameters correspond to the optimal operating conditions. 
}
\label{fig5}
\end{center}
\end{figure}
%%%%%%%

It is instructive to look at the scattering amplitude $S\left( t,E \right)$ as a function of its arguments, Fig.~\ref{fig5}. 
The narrow chine visualizes a quantum level in the cavity moving under the action of the potential $U(t)$. 
At zero temperature the scattering amplitude  at the Fermi energy, $E = \mu$, is sufficient to calculate the emitted current: $I_{a}(t) = -ie/(2\pi) S \partial S^{*}/\partial t$. \cite{Buttiker:1994vl,Brouwer:1998jt,Avron:2001fl} 
Thus the cross-section on Fig.~\ref{fig5} at $E=0$ shows us when a quantum level crosses the Fermi energy and, hence, when the current pulses appear. 
Importantly, the shape of peaks of the aforementioned  cross-section (at $T\ll 1$) is similar to the shape of a current pulse (up to a normalization factor). 
This can be easily shown if one considers the scattering amplitude close to, say, the time of an electron emission, $t_{-}$. 
It reads:\cite{OSMB08} 

\begin{eqnarray}
S(t,\mu) = (t - t_{-} + i \Gamma_{\tau})/(t - t_{-} - i \Gamma_{\tau}) \,.
\label{sfroz}
\end{eqnarray}
\ \\
\noindent
Then the current $I_{a}(t)$, Eq.~(\ref{pulse-a}), is expressed in terms of the real part of the scattering amplitude as follows,

\begin{eqnarray}
I_{a}(t) = \frac{e }{2\pi \Gamma_{\tau} } \left[ 1 - {\rm Re} S(t,\mu) \right]\,.
\label{cvscoh}
\end{eqnarray}
\ \\
\noindent

\subsection{Non-adiabatic emission}

The periodic pulsed potential [$U(t) = U(t + {\cal T})$], 

\begin{equation}
U(t) = \left\{
\begin{array}{ll}
U_{0} \,,  &-{\cal T}/2 < t <  t_{-}\,,\\
 &  \\
U_{1} \,,  &t_{-} < t <  t_{+}\,,\\
 &  \\
U_{0} \,,  & t_{+} < t <  {\cal T}/2\,,\\
 &  \\
\end{array}
\right.
\label{pulse} 
\end{equation}
\ \\
\noindent
is an example, relevant to experiment, \cite{Feve:2007jx} leading to a non-adiabatic emission. 
The non-adiabatic behavior is caused by the potential jumps, which formally have to be sharp on the scale of $\tau$. 
Before calculating the scattering amplitude for the pulsed potential  $U(t)$, Eq.~(\ref{pulse}), we consider the following auxiliary problem. 

\subsubsection{Single-step potential}

Let us find scattering amplitudes for a cavity driven by the single-step potential,

\begin{equation}
U(t) = \left\{
\begin{array}{ll}
U_{0} \,,  &t <  0\,,\\
 &  \\
U_{1}\,, &  t > 0 \,.
\end{array}
\right.
\label{step} 
\end{equation}
\ \\
\noindent
With this potential the time-dependent phase, say, $\Phi_{in,q}(t)$, Eq.~(\ref{phi-in}), can be easily calculated:

\begin{equation}
\Phi_{in,q}(t) = \left\{
\begin{array}{ll}
2\pi q \frac{eU_{1}}{h/\tau} \,, &  t >  q\tau \,,\\
\ \\
2\pi  \frac{t}{\tau} \frac{e \delta U  }{h/\tau} + 2\pi q \frac{eU_{0}}{h/\tau} \,, & 0 <  t < q\tau \,, \\
\ \\
2\pi q \frac{eU_{0}}{h/\tau} \,, & t < 0\,.\\
\end{array}
\right.
\label{phi-in-step} 
\end{equation}
\ \\
\noindent
Here $\delta U = U_{1} - U_{0}$.
To sum up over $q$ in Eq.~(\ref{in}) we note that for a given $t < 0$ we have to use $U_{0}$ $\forall q$.
In contrast, as far as $ N\tau < t <  (N+1)\tau$ we have to use $U_{1}$ for $q \leq N$ and a more complicated phase  for $q > N$. 
We can represent a time-dependent scattering amplitude as follows:

\begin{subequations}
\label{instep}
\begin{eqnarray}
S_{in}(t,E) = S(U[t];E) + \theta(t) \delta S_{in}(t,E) \,.
\label{in-step}
\end{eqnarray}
\ \\
\noindent
Here $\theta(t)$ is the Heaviside theta-function, $\theta(t) = 0$ for $t<0$ and $\theta(t) = 1$ for $t > 0$. 
The frozen amplitude $S(U[t];E)$ is given by Eq.~(\ref{froz}), and $\delta S_{in}$ is given within each interval $N\tau < t <  (N+1)\tau$  ($N = 0,1,2,\dots$)  as

\begin{eqnarray}
\delta S_{in}(t,E) = e^{ i\theta_{r} }\, T\, R^{\frac{N}{2} }   \nonumber \\
\label{d-in-step} \\
 \times \left\{\frac{ e^{i (N+1) \phi_{1}(E)} }{1 - \sqrt{R} e^{i\phi_{1}(E) } } -  e^{- i 2\pi  \frac{t}{\tau} \frac{e\delta U}{h/\tau} } \frac{ e^{i (N+1) \phi_{0}(E)} }{1 - \sqrt{R} e^{i\phi_{0}(E) } }  \right\}  , \nonumber 
\end{eqnarray}
\end{subequations}
\ \\
\noindent
with $T = 1 - R$ a transmission coefficient and $\phi_{k}(E) =  \phi(E;U_{k})$, $k = 0,\, 1$. 

The quantity $\delta S_{in}(t,E)$ characterizes how $S_{in}(t,E)$ deviates from the stationary scattering amplitude $S(U;E)$ corresponding to the instantaneous potential $U= U(t)$. 
This deviation exists only after the potential was changed, $t > 0$, and it decreases,  

\begin{eqnarray}
\delta S_{in} \sim e^{- \frac{t}{2\tau_{D} } }\,, \quad t \gg \tau \,,
\label{decay}
\end{eqnarray}
\ \\
\noindent
with a characteristic time 

\begin{eqnarray}
\tau_{D} = \tau/\ln(1/R) \,.
\label{taud-1}
\end{eqnarray}

An analogous calculation gives 

\begin{subequations}
\label{outstep}
\begin{eqnarray}
S_{out}(E,t) = S(U[t];E) + \theta(-t) \delta S_{out}(E,t) \,.
\label{out-step}
\end{eqnarray}
\ \\
\noindent
where within each interval $-(N+1)\tau < t < - N\tau$, 

\begin{eqnarray}
\delta S_{out}(E,t) = e^{ i\theta_{r} }\, T\, R^{\frac{N}{2} } \nonumber \\
\label{d-out-step} \\
\times\left\{ \frac{ e^{i (N+1) \phi_{0}(E) } }{1 - \sqrt{R} e^{i\phi_{0}(E) } }  -  e^{- i 2\pi  \frac{t}{\tau} \frac{e\delta U}{h/\tau} } \frac{ e^{i (N+1) \phi_{1}(E) } }{1 - \sqrt{R} e^{i\phi_{1}(E) } } \right\}  . \nonumber
\end{eqnarray}
\end{subequations}
\ \\
\noindent
In contrast to $S_{in}$ the scattering amplitude $S_{out}(E,t)$ deviates from the frozen scattering amplitude, $S(E;U[t])$, at times preceding the change of a potential. 
At $|t| \gg \tau$ the deviation $\delta S_{out}$ decays exponentially with a characteristic time $\tau_{D}$.

\subsubsection{Optimal operating conditions}

The calculations are simplified greatly for the  optimal operating conditions\cite{Feve:2007jx,Parmentier11} which lead to the emission of a single electron and hole during each period. One condition is that the potential changes by exactly one level spacing $\Delta = h/\tau$,

\begin{subequations}
\label{op}
\begin{eqnarray}
e\delta U = - \chi\, \Delta \,,
\label{opt}
\end{eqnarray}
\ \\
\noindent
where $\chi = \mp $. 
In addition the Fermi energy should lie exactly in the middle of two neighboring quantum levels of the cavity,

\begin{eqnarray}
\theta_{r} + \varphi(\mu) = \pi\,.
\label{neighbor}
\end{eqnarray}
\end{subequations}
\ \\ 
\noindent
With these conditions the frozen amplitude becomes independent of time,  $S(E) \equiv S(U_{0};E) = S(U_{1};E)$. 
In other words, mere shaking of a potential would not disturb an electron system. 
What causes a dynamical (transient) response is electrons entering and leaving the cavity at different potentials. 
That is described by $\delta S_{\chi}(E,t) \equiv \delta S_{in}(t,E) = \delta S_{out}(E,t)$, 

\begin{subequations}
\label{dstep}
\begin{eqnarray}
\delta S_{\chi}(E,t) = e^{ i\theta_{r} } T\,    \frac{ R^{\frac{N}{2} } e^{i (N+1) \phi(E) } \left(   1 -  e^{\chi i 2\pi  \frac{t}{\tau}  } \right) }{1 - \sqrt{R} e^{i\phi(E) } } \,, \nonumber \\
\label{d-step} 
\end{eqnarray}
\ \\
\noindent
with 

\begin{eqnarray}
\phi(E) = \pi + 2\pi \frac{ E - \mu }{ \Delta}    \,.
\label{phi-t-1}
\end{eqnarray}
\end{subequations}
For simplicity all calculations from here on are done for optimal operating conditions. 
However the formalism used in itself is not restricted to the optimal conditions, Eqs.~(\ref{opt}) and (\ref{neighbor}). 

%%%%%%%
\begin{figure}[b]
\begin{center}
\includegraphics[width=86mm]{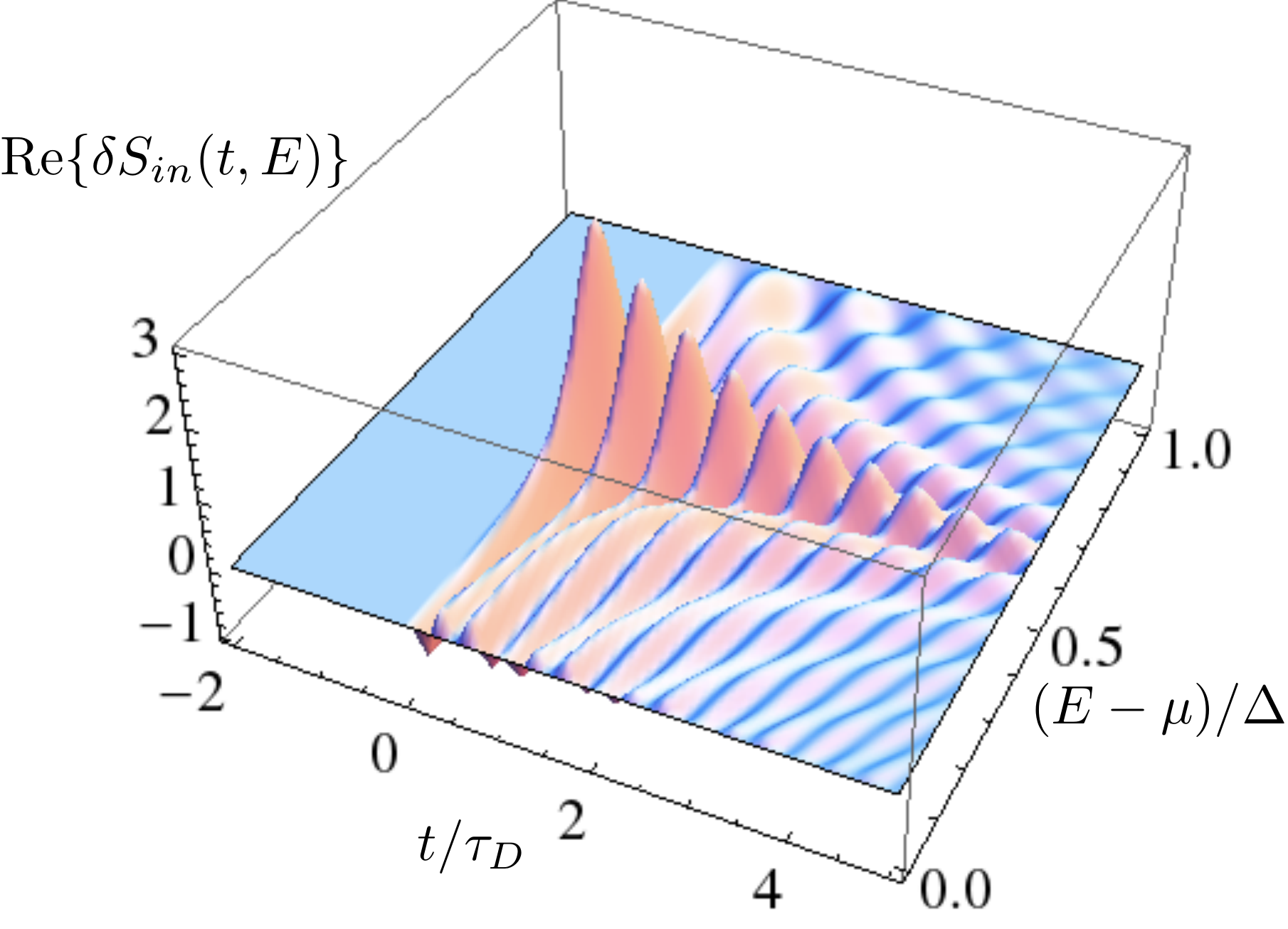}
\caption{
(Color online) Non-adiabatic emission: The real part of $\delta S_{in}\left( t,E \right)$, Eq.~(\ref{d-step}) with $\chi = -$, is shown close to the time of emission of an electron, $t \sim t_{-}$ . The time $t$ is measured in units of the dwell time $\tau_{D} = \tau/T$. The energy $E$ is measured from the Fermi energy $\mu$ in units of the level spacing $\Delta$. Only one period for $E$ is shown. The visible ripples reflect oscillations in time with the period of $\tau$.  The transmission probability of the quantum point contact connecting a cavity is $T=0.5$. Other parameters correspond to the optimal operating conditions. 
}
\label{fig6}
\end{center}
\end{figure}
%%%%%%%

The real part of the scattering amplitude $\delta S_{\chi}$, Eq.~(\ref{d-step}), is shown in Fig.~\ref{fig6}. 
Its overall behavior in time reflects the asymmetry of the emitted state, in particular, of the current pulse $I_{na}\left( t \right)$, Eq.~(\ref{pulse-na}). 
In addition it illustrates that the largest variations of the scattering amplitude occur at the energy of the quantum state in the cavity.

\subsubsection{Pulsed potential} 

Now we come back to the periodic pulsed potential, Eq.~(\ref{pulse}). 
For definiteness we use $eU_{1} > eU_{0}$. 
We suppose a drive with a delay between the potential steps that is long compared to the difference between absorption and emission times which in turn are taken to be long compared to the dwell time,

\begin{eqnarray}
{\cal T}  > t_{+} - t_{-} \gg \tau_{D} \,. 
\label{long}
\end{eqnarray}
\ \\
\noindent
Therefore the transient behavior caused by one potential step vanishes completely before the next step appears. This permits us to  use the results for a single-step potential and get ($t \in [-{\cal T}/2, {\cal T}/2] $)

\begin{eqnarray}
S_{in}(t,E) &=& S(E) + \theta(t-t_{-}) \delta S_{-}(E,t- t_{-}) \nonumber \\
&&+ \theta(t-t_{+}) \delta S_{+}(E,t- t_{+}) \,, \nonumber \\
\label{double} \\
S_{out}(E,t) &=& S(E) + \theta(t_{-} - t) \delta S_{-}(E,t-t_{-}) \nonumber \\ 
&& + \theta(t_{+} - t) \delta S_{+}(E,t-t_{+}) \,. \nonumber 
\end{eqnarray}
\ \\
\noindent
Note that at the time $t_{-}$ an electron is emitted by the driven cavity whereas at the time $t_{+}$ a hole is emitted.

\subsubsection{Fourier coefficients}

To calculate the Floquet scattering amplitudes, see Eqs.~(\ref{fl}),  we need the Fourier transformation of Eq.~(\ref{double}). 
To integrate over time we use the following trick: Since $\delta S_{\chi}$ is constant over an  interval of duration $\tau$, we integrate over this interval the factor $\exp(i n \Omega t)$ only and then sum up over $N$.  
Under the condition of Eq.~(\ref{long},) the sum over $N$ runs from $0$ to $\infty$.
As a result we find:   

\begin{subequations}
\label{inn}
\begin{eqnarray}
S_{in,n}(E) &=& S(E)  \delta_{n,0} - A_{n}(E) \left\{  \frac{ e^{i n \Omega t_{-}} }{1 -  \frac{n\hbar\Omega }{\Delta }  } + \frac{ e^{i n \Omega t_{+}} }{1 +  \frac{n\hbar\Omega }{\Delta } }  \right\}  , \nonumber \\
\label{in-n} \\
S_{out,n}(E) &=& S_{in,n}(E- n\hbar\Omega) \,. \nonumber 
\end{eqnarray}
\ \\
\noindent
Here 

\begin{eqnarray}
A_{n}(E) =  S(E)  e^{ i \pi \frac{n \hbar\Omega  }{\Delta }  }\,\frac{\sin\left( \pi \frac{n \hbar\Omega }{\Delta } \right) }{\pi n   }  \nonumber \\ 
\label{a} \\
\times \frac{ T }{\left( 1 - \sqrt{R} e^{-i\phi(E)}  \right) \left( 1 - \sqrt{R} e^{i(\phi(E) + n\Omega\tau  )}  \right) }  \,, \nonumber
\end{eqnarray} 
\end{subequations}
\ \\
\noindent
and $\delta_{n,0}$ is the Kronecker symbol.

\subsubsection{Continuous frequency representation}

From Eq.~(\ref{long}) it follows that $\hbar\Omega \ll \Delta$. 
Therefore, there are many ($n \sim \Delta/(\hbar\Omega) \gg 1$) photon-assisted amplitudes contributing to scattering. 
Since the replacement $n\to n+1$ changes the scattering amplitude only a little, it is convenient to go over from the discrete frequency representation to the continuous frequency representation. 
For this purpose we use the following correspondence,

\begin{subequations}
\label{dimless}
\begin{eqnarray}
n\Omega \to \Omega_{n} \,, &\quad& \sum\limits_{n = - \infty}^{\infty} \to \int\limits_{-\infty}^{\infty} \frac{d\Omega_{n} }{\Omega } \,, \nonumber \\
\label{corr} \\
\delta_{n,0} \to \Omega \delta(\Omega_{n})\,, &\quad& \int\limits_{0}^{\cal T} dt^{\prime}  e^{in\Omega t^{\prime}} \to \int\limits_{-\infty}^{\infty} dt^{\prime}  e^{i\Omega_{n} t^{\prime}  } \,. \nonumber 
\end{eqnarray}
\ \\
\noindent 
Here $\delta(\Omega_{n})$ is the Dirac delta-function. 

To simplify long equations we also introduce the following dimensionless quantities,

\begin{eqnarray}
\epsilon = \frac{E - \mu }{\Delta }\,, \quad \omega_{n} = \frac{\hbar\Omega_{n} }{\Delta } \,,
\label{dimless-a}
\end{eqnarray}
\ \\
\noindent
and the abbreviation,

\begin{eqnarray}
\rho(\epsilon) = \frac{ 1 - \sqrt{R} e^{i \phi(\epsilon)} }{\sqrt{T} } \,, 
\label{dimles-b}
\end{eqnarray}
\end{subequations}
\ \\
\noindent
where $\phi(\epsilon) = \pi + 2\pi \epsilon$, see Eq.~(\ref{phi-t-1}). 
Note the density of states, $\nu(E) = (i/{2\pi}) S(E) \partial S^{*}(E)/\partial E$, \cite{Buttiker:1994vl} of the cavity can be written as $\nu(E) \Delta = 1/|\rho(E)|^{2}$.

With this  Eqs.~(\ref{in-n}), originally expressed in as a discrete Fourier transformation, now take the form of a continuous Fourier transformation, 

\begin{subequations}
\label{n-1}
\begin{eqnarray}
S_{in}(\omega_{n}, \epsilon) &=& S(\epsilon) \frac{\hbar\Omega }{\Delta }  \frac{\sin(\pi \omega_{n}) }{\pi \omega_{n} } \, e^{i \pi \omega_{n}}  
\label{in-n-1} \\
&&\times \left\{  \delta(\omega_{n}) - \frac{ \frac{ e^{i 2\pi \omega_{n} \frac{t_{-} }{\tau } }}{1 -  \omega_{n}  } + \frac{ e^{i 2\pi \omega_{n} \frac{t_{+} }{\tau } }}{1 + \omega_{n} } }{\rho^{*}(\epsilon) \rho(\epsilon + \omega_{n}) }   \right\} , \nonumber  
\end{eqnarray}
\begin{eqnarray}
S_{out}(\epsilon,\omega_{n}) &=& S(\epsilon) \frac{\hbar\Omega }{\Delta }  \frac{\sin(\pi \omega_{n}) }{\pi \omega_{n} } \, e^{- i \pi \omega_{n}}  
\label{out-n-1} \\
&&\times \left\{  \delta(\omega_{n}) - \frac{ \frac{ e^{i 2\pi \omega_{n} \frac{t_{-} }{\tau } }}{1 -  \omega_{n}  } + \frac{ e^{i 2\pi \omega_{n} \frac{t_{+} }{\tau } }}{1 + \omega_{n} } }{\rho^{*}(\epsilon) \rho(\epsilon - \omega_{n}) }   \right\} . \nonumber 
\end{eqnarray}
\end{subequations}
\ \\
\noindent
Note here we used the following  property of the Dirac delta-function: $\delta(\omega_{n}\Delta /\hbar) = (\hbar/\Delta) \delta(\omega_{n})$.

\subsubsection{Unitarity}

It is instructive to verify that the Floquet scattering matrix we calculated is unitary. 
Let us, for instance, prove the following unitarity condition:\cite{Moskalets:2011cw}

\begin{eqnarray}
\sum\limits_{p = -\infty}^{\infty} S_{F}^{*}(E + p\hbar\Omega, E - m\hbar\Omega) \nonumber \\
\label{uni} \\
\times S_{F}(E + p\hbar\Omega, E - n\hbar\Omega) = \delta_{m,n} \,, \nonumber
\end{eqnarray}
\ \\
\noindent
where $p,\, m,\, n$ all are integers. 
Using Eq.~(\ref{flb}) and the normalized quantities of Eq.~(\ref{dimless-a}) and going over to the continuous frequency representation we arrive at the following identity to prove:  

\begin{eqnarray}
\int\limits_{-\infty}^{\infty} d\omega_{p} S_{out}^{*}(\epsilon + \omega_{p}, \omega_{m} + \omega_{p}) S_{out}(\epsilon + \omega_{p}, \omega_{n} + \omega_{p}) \nonumber \\
= \left( \frac{\hbar\Omega }{\Delta } \right)^{2} \delta(\omega_{m} - \omega_{n})\,. \, \quad \, \quad \, \label{uni-1}
\end{eqnarray}
\ \\
\noindent
With Eq.~(\ref{in-n-1}) we get

\begin{eqnarray}
\int\limits_{-\infty}^{\infty}d\omega_{p} \frac{\sin(\pi [ \omega_{m} + \omega_{p}] ) \sin(\pi [ \omega_{n} + \omega_{p}])}{\pi^{2} [ \omega_{m} + \omega_{p}] [ \omega_{n} + \omega_{p}] }  e^{i \pi [\omega_{m} - \omega_{n}]} \nonumber \\
\label{uni-1-2} \\
\times \left\{  \delta(\omega_{m} + \omega_{p}) - \frac{ \frac{ e^{-i 2\pi [\omega_{m} + \omega_{p}] \frac{t_{-} }{\tau } }}{1 -  \omega_{m} - \omega_{p}  } + \frac{ e^{-i 2\pi [\omega_{m} + \omega_{p}] \frac{t_{+} }{\tau } }}{1 + \omega_{m} + \omega_{p} } }{\rho(\epsilon+\omega_{p}) \rho^{*}(\epsilon - \omega_{m}) }   \right\} \nonumber \\
\times \left\{  \delta(\omega_{n} + \omega_{p}) - \frac{ \frac{ e^{i 2\pi [\omega_{n} + \omega_{p}] \frac{t_{-} }{\tau } }}{1 -  \omega_{n} - \omega_{p}  } + \frac{ e^{i 2\pi [\omega_{n} + \omega_{p}] \frac{t_{+} }{\tau } }}{1 + \omega_{n} + \omega_{p} } }{\rho^{*}(\epsilon+\omega_{p}) \rho(\epsilon - \omega_{n}) }   \right\} \nonumber \\
= \delta(\omega_{m} - \omega_{n}) \,. \nonumber 
\end{eqnarray}
\ \\
\noindent
Here we used the property of the stationary scattering amplitude: $|S(\epsilon + \omega_{p})|^{2} = 1$.
Next we open the curly brackets, 

\begin{eqnarray}
\int\limits_{-\infty}^{\infty} \frac{ d\omega_{p} }{|\rho(\epsilon+\omega_{p})|^{2} }  \frac{\sin(\pi [ \omega_{m} + \omega_{p}] ) \sin(\pi [ \omega_{n} + \omega_{p}])}{\pi^{2} [ \omega_{m} + \omega_{p}] [ \omega_{n} + \omega_{p}] }  \nonumber \\
\label{uni-1-3} \\
\times \bigg\{ \xi_{p} +  \frac{e^{ -i  2\pi [ \omega_{m} - \omega_{n}] \frac{ t_{-} }{\tau }  }  }{ \left( \omega_{m} + \omega_{p}  -1 \right)\left( \omega_{n} + \omega_{p}  -1 \right) } \nonumber \\
 + \frac{e^{ -i 2\pi  [ \omega_{m} - \omega_{n}] \frac{ t_{+} }{\tau }  } }{ \left( \omega_{m} + \omega_{p}  +1 \right)\left( \omega_{n} + \omega_{p}  +1 \right) }  \bigg\} =  \frac{ 2 \sin(\pi  [\omega_{m} - \omega_{n}] )  }{\pi [\omega_{m} - \omega_{n}]  }  \nonumber \\
\nonumber \\
\times  \frac{  e^{ -i 2\pi  [ \omega_{m} - \omega_{n}] \frac{ t_{-} }{\tau }  } + e^{ -i 2\pi [ \omega_{m} - \omega_{n}] \frac{ t_{+} }{\tau }  }   }{1 -  [\omega_{m} - \omega_{n}]^{2}    }  \nonumber \,,
\end{eqnarray}
\ \\
\noindent
where

\begin{eqnarray}
\xi_{p} = - e^{-i 2\pi \omega_{p} \frac{ t_{+} - t_{-} }{\tau } }  \frac{ e^{ i  2\pi  \omega_{n} \frac{ t_{-} }{\tau }  }\, e^{ -i  2\pi  \omega_{m} \frac{ t_{+} }{\tau }  }  }{ \left( \omega_{n} + \omega_{p}  -1 \right)\left( \omega_{m} + \omega_{p}  +1 \right) } \nonumber \\
\label{xi} \\
- e^{i 2\pi \omega_{p} \frac{ t_{+} - t_{-} }{\tau } }  \frac{ e^{ i 2\pi  \omega_{n} \frac{ t_{+} }{\tau }  }\, e^{ -i 2\pi  \omega_{m} \frac{ t_{-} }{\tau }  } }{ \left( \omega_{n} + \omega_{p}  +1 \right)\left( \omega_{m} + \omega_{p}  -1 \right) }  \,. \nonumber
\end{eqnarray}
\ \\
\noindent
Since the time period between the potential steps is much larger than the duration of one turn, $t_{+} - t_{-} \gg \tau$, see Eqs.~(\ref{long}) and (\ref{taud}), the quantity $\xi_{p}$ oscillates fast as a function of $\omega_{p}$. 
The terms under the integral over $\omega_{p}$ which are a product of a function that oscillates fast with a smooth function are zero. 
Hence we can ignore $\xi_{p}$ in Eq.~(\ref{uni-1-3}).
Physically it means that the emission of an electron at time $t_{-}$ has no any effect on the emission of a hole at time $t_{+}$. 
Therefore, one can calculate quantities (current, heat, etc.) caused separately by either electrons or holes.
To this end in Eqs.~(\ref{n-1}) we remove the part with either $e^{i 2\pi \omega_{n} \frac{t_{+} }{\tau } }$ or $e^{i 2\pi \omega_{n} \frac{t_{-} }{\tau } }$, respectively. 

To prove Eq.~(\ref{uni-1-3}) (without $\xi_{p}$) we note that $t_{-}$ and $t_{+}$ are arbitrary and, therefore, the parts with the factors $e^{ - i 2\pi  [\omega_{m} - \omega_{n}] \frac{ t_{-} }{\tau }  }$ or $e^{ -i 2\pi  [\omega_{m} - \omega_{n}] \frac{ t_{+} }{\tau }  }$ have to be considered separately. 
Therefore we have to show that,

\begin{eqnarray}
\int\limits_{-\infty}^{\infty} \frac{ d\omega_{p} }{|\rho(\epsilon+\omega_{p})|^{2} }  \frac{\sin(\pi [ \omega_{m} + \omega_{p}] ) \sin(\pi [ \omega_{n} + \omega_{p}])}{\pi^{2} [ \omega_{m} + \omega_{p}] [ \omega_{n} + \omega_{p}] }  \nonumber \\
\label{uni-1-4} \\
\times   \frac{1  }{ \left( \omega_{m} + \omega_{p}  \mp 1 \right)\left( \omega_{n} + \omega_{p}  \mp 1 \right) } =  \frac{ 2 \sin(\pi   \omega_{q} )  }{\pi \omega_{q} \left( 1 -  \omega_{q}^{2} \right) }   \nonumber \,,
\end{eqnarray}
\ \\
\noindent
where $\omega_{q} = \omega_{m} - \omega_{n}$. 
To simplify calculations we do the following: 
Since $\rho(\epsilon + \omega_{p})$, Eq.~(\ref{dimles-b}), is periodic in $\omega_{p}$ with period $1$, we integrate over one period and sum up contributions from all periods.
So we replace,

\begin{eqnarray}
\int\limits_{-\infty}^{\infty} d\omega_{p} &\to& \sum\limits_{a = -\infty}^{\infty}\, \int\limits_{0}^{1} d\omega_{p}^{\prime} \,, \nonumber \\
\label{int} \\
\omega_{p} &\to& \omega_{p}^{\prime} + a \,, \nonumber
\end{eqnarray}
\ \\
\noindent  
and get

\begin{eqnarray}
\int\limits_{0}^{1} d\omega_{p}^{\prime}\, \frac{ \Sigma_{q}  }{|\rho(\epsilon+\omega_{p}^{\prime})|^{2} }  =  \frac{ 2 \sin(\pi   \omega_{q} )  }{\pi \omega_{q} \left( 1 -  \omega_{q}^{2} \right) } \,,
\label{uni-1-5} 
\end{eqnarray}
\ \\
\noindent
where 

\begin{eqnarray}
\Sigma_{q} =  \sum\limits_{a = -\infty}^{\infty}  \frac{\sin(\pi [ \omega_{m} + \omega_{p}^{\prime}] ) \sin(\pi [ \omega_{n} + \omega_{p}^{\prime}])}{\pi^{2} [ \omega_{m} + \omega_{p}^{\prime} + a] [ \omega_{n} + \omega_{p}^{\prime} + a] }  \nonumber \\
\label{sigq} \\
\times   \frac{1  }{ \left( \omega_{m} + \omega_{p}^{\prime} + a  \mp 1 \right)\left( \omega_{n} + \omega_{p}^{\prime} + a  \mp 1 \right) } \nonumber\,.
\end{eqnarray}
\ \\
\noindent
To calculate $\Sigma_{q}$ we use the following identity,

\begin{eqnarray}
\sigma_{2} \equiv  \sum\limits_{a=-\infty}^{\infty} \frac{1 }{\left\{(a  - \delta)^{2} - \frac{1}{4} \right\} \left\{ \left(a  - [x+\delta] \right)^{2} - \frac{1}{4} \right\}} \nonumber \\
\label{sum2} \\
=    \frac{\sin(\pi x)  }{ x \left( 1-  x^{2} \right)} \frac{2\pi }{\cos(\pi\delta) \cos(\pi[x+ \delta])  } \,, \nonumber 
\end{eqnarray}
\ \\
\noindent
which can be proven with the help of the following text-book sum  

\begin{eqnarray}
\sigma_{0}(\gamma) &\equiv&  \sum\limits_{a=-\infty}^{\infty} \frac{1 }{a + \gamma } = \pi \cot(\pi \gamma)  \,, 
\nonumber
\end{eqnarray}
\ \\
\noindent
taken with different arguments:

\begin{eqnarray}
\sigma_{2} &=& \left\{ \sigma_{0}\left(-[x+\delta] \pm \frac{1}{2} \right) -  \sigma_{0}\left(-\delta \pm \frac{1}{2} \right)   \right\}  \nonumber \\
&&\times \left\{ \frac{2}{x} + \frac{1}{1 - x}  - \frac{1}{1+x}\right\} \,.
\nonumber
\end{eqnarray}

So, in Eq.~(\ref{sigq}) we introduce $-\delta \pm 0.5 = \omega_{m} + \omega_{p}^{\prime}$ and $- [x+\delta] \pm 0.5 = \omega_{n} + \omega_{p}^{\prime}$, use Eq.~(\ref{sum2}), and obtain

\begin{eqnarray}
\Sigma_{q} =  \frac{2\sin(\pi \omega_{q})  }{ \pi \omega_{q} \left( 1-  \omega_{q}^{2} \right)} \,.
\label{sigq-1}
\end{eqnarray}
\ \\
\noindent
Since $\Sigma_{q}$ is independent of $\omega_{p}^{\prime}$, we can integrate in Eq.~(\ref{uni-1-5}). 
With $\rho$ given in Eq.~(\ref{dimles-b}) we get one. 
Therefore, the use of Eq.~(\ref{sigq-1}) in Eq.~(\ref{uni-1-5}) gives identity.
This completes the proof of Eq.~(\ref{uni-1}).

\section{Zero-frequency noise power}
\label{shn}

Consider the conductor shown in Fig.~\ref{fig3} as a four-probe conductor with seperate contacts $1, 2, 3$ and $4$. For this four probe conductor the zero-frequency correlation function \cite{Blanter:2000wi}, ${\cal P}_{34}$, of currents flowing into these contacts is \cite{MBnoise04} 

\begin{eqnarray}
{\cal P}_{34 }  &=& \frac{{e^2 }}{h} \int\limits_0^\infty  dE\, \sum\limits_{\gamma  = 1}^{2 } \sum\limits_{\delta  = 1}^{2 } \sum\limits_{n,m=- \infty }^\infty    \frac{{\left[f \left( {E_n } \right) - f \left( {E_m } \right) \right]^2 }}{2} \nonumber \\
&&  \times \sum\limits_{p=- \infty }^\infty  S_{F,3 \gamma }^* \left( {E, E_n } \right)S_{F,3 \delta } \left( {E, E_m } \right) 
\label{noise-gen} \\
\nonumber \\
&&\times S_{F,4 \delta }^* \left( {E_p, E_m } \right)S_{F,4 \gamma } \left( {E_p, E_n } \right) . \nonumber
\end{eqnarray}
\ \\
\noindent
Here $f(E)$ is the Fermi distribution function, the same for both contacts $1$ and $2$, $E_{n} = E + n\hbar\Omega$. 
The elements of the Floquet scattering matrix of the circuit, $\hat S_{F}$, are expressed in terms of the Floquet scattering amplitudes of the sources $A$ and $B$. 
For example,  

\begin{eqnarray}
S_{F,31}(E_{p}, E_{n}) &=& t_{C}\,  e^{i \varphi_{L^{A}}(E)} e^{i p \Omega \tau_{L^{A}} } S^{A}_{out,p-n}(E_{p}) \,, \nonumber \\
\label{elements} \\
S_{F,32}(E_{p}, E_{n}) &=& r_{C}\,  e^{i \varphi_{L^{B}}(E)} e^{i p \Omega \tau_{L^{B}} } S^{B}_{out,p-n}(E_{p}) \,, \nonumber 
\end{eqnarray}
\ \\
\noindent
where $r_{C}$/$t_{C}$  is the reflection/transmission amplitude at the QPC C assumed to be energy independent, $L^{j}$ the distance to the cavity $j = A,\, B$ from the QPC C, $\varphi_{L^{j}}(E)$ the phase factor corresponding to free propagation from the cavity $j$ to the QPC C, $\tau_{L^{j}}$ the time of flight from the cavity $j$ to the QPC C. 
We assume linear dispersion for free electrons and, therefore, use $\varphi_{L^{j}}(E_{p}) = \varphi_{L^{j}}(E) + p\Omega \tau_{L^{j}}$.
We dropped unimportant phase factors related to free propagation from the QPC C to the metallic contacts.

\subsection{Quantized noise of a single source}
\label{shn-1}

For a moment we switch of, say, the source B. 
Now we use $S^{B}_{out,p-n}(E_{p}) = \delta_{p,n}$, in Eq.~(\ref{elements}) and reduce Eq.~(\ref{noise-gen}) to

\begin{eqnarray}
{\cal P}_{34 }  &=& -\,\frac{{e^2 }}{h} (1-T_{C}) T_{C} \int\limits_0^\infty  dE\, \sum\limits_{n= - \infty }^\infty    \left\{f \left( {E_{-n} } \right) - f \left( {E } \right) \right\}^2   \nonumber \\
&& \times  \left | S_{out,n}^{A} \left( E \right)  \right |^{2} .
\label{noise-1} 
\end{eqnarray}
\ \\
\noindent
Here we changed $n \to - n$ and used both the following relation $r_{C}t_{C}^{*} = - r_{C}^{*}t_{C}$  and the unitarity condition for $S^{A}_{out}$,

\begin{eqnarray}
\sum\limits_{p=-\infty}^{\infty} S^{A*}_{out,p-m}(E_{p}) S_{out,p-n}^{A}(E_{p}) = \delta_{n,m} \,,
\label{uniF}
\end{eqnarray}
\ \\
\noindent
which follows directly from Eq.~(\ref{uni}). 

Next with the quantities introduced in Eqs.~(\ref{dimless}) and with Eq.~(\ref{out-n-1}) we rewrite Eq.~(\ref{noise-1}) as follows:

\begin{eqnarray}
{\cal P}_{34 }^{na}  &=& -\,{\cal P}_{0}  \int\limits_{-\infty}^\infty  d\epsilon\int\limits_{-\infty}^\infty  d\omega_{n}\,   \frac{ \left\{f \left( \epsilon - \omega_{n}  \right) - f \left( \epsilon \right) \right\}^2  }{|\rho(\epsilon - \omega_{n} )|^{2} |\rho(\epsilon )|^{2}}\nonumber \\
&& \times \frac{  \sin^{2}(\pi  \omega_{n})  }{\pi^{2} \omega_{n}^{2}  } \left\{ \frac{ 1 }{\left( 1 - \omega_{n}  \right)^{2} } +  \frac{   1 }{\left(1 + \omega_{n}  \right)^{2} }  \right\}  .
\label{noise-1-1} 
\end{eqnarray}
\ \\
\noindent
Here we have dropped the terms $\sim \exp( i 2\pi [\epsilon + \omega_{n}] (t_{-}^{A} - t_{+}^{A} )/\tau )$ as non-contributing.
Since $(t_{-}^{A} - t_{+}^{A} )/\tau \gg 1$ these terms oscillate fast in both $\epsilon$ and $\omega_{n}$. 
Therefore, they are nullified after the integration. 
The upper index ``$na$'' emphasizes that this equation is for the time-dependent potential $U(t)$, Eq.~(\ref{pulse}), leading to non-adiabatic emission of particles.  
Note the parts proportional $1/(1 - \omega_{n} )^{2}$ and $1/(1 + \omega_{n})^{2}$  correspond  to an electron and a hole contributions, respectively.

With $\rho(\epsilon)$, Eq.~(\ref{dimles-b}), in the limit of $T\to 0$ we represent the density of states (normalized to $\Delta$) as a sum of Breit-Wigner resonances each of unit area: 

\begin{eqnarray}
\frac{1}{|\rho(\epsilon)|^{2}} &=& \sum\limits_{\ell = -\infty}^{\infty} \frac{g/\pi }{(\epsilon + 0.5 - \ell)^{2} + g^{2} }\,, \nonumber \\
\label{BW} \\
\frac{1}{|\rho(\epsilon - \omega_{n})|^{2}} &=& \sum\limits_{n = -\infty}^{\infty} \frac{g/\pi }{(\epsilon  - \omega_{n} + 0.5 - n)^{2} + g^{2} }\,, \nonumber 
\end{eqnarray}
\ \\
\noindent
with $g = T/(4\pi)$ a width (normalized to $\Delta$) .

Integrating over $\omega_{n}$, we take into account that the integrand has narrow peaks at the integers $\omega_{n} = \ell - n$, where the sinus is zero. 
Therefore, to leading order in $g \ll 1$ what matters is only $\ell = n$, $\ell = n \pm 1$ when the zeros of the denominator cancel the zero of $\sin(\pi \omega_{n})$.
Because of the difference of the Fermi distribution functions, the term with $\ell = n$ does not contribute.
In addition, if the temperature is much lower then the level spacing, we can approximate $f(\epsilon) \approx \theta(-\epsilon)$.
Using this, we find that only the pairs $\ell = 1\,, n = 0$ (an electron emission) and  $\ell = 0\,, n = 1$ (a hole emission) contribute. 
Therefore, we arrive at

\begin{eqnarray}
{\cal P}_{34}^{na} = - 2 {\cal P}_{0} \,,
\label{noise-1-2}
\end{eqnarray}
\ \\
\noindent
announced already in Eq.~(\ref{sh-1}).

\subsection{Shot noise suppression effect}
\label{shn-2}

If both sources A and B are switched on then  

\begin{eqnarray}
{\cal P}_{34 }  = -\, {\cal P}_{0}  \frac{\Delta^{4}}{(\hbar\Omega)^{4}}  \int\limits_{-\infty}^\infty  d\epsilon\,   \int\limits_{-\infty}^\infty  d\omega_{m}\, \int\limits_{-\infty}^\infty  d\omega_{n}\,       \nonumber \\
\{f \left( \epsilon + \omega_{n}  \right)  - f \left( \epsilon + \omega_{m} \right) \}^2 \ 
\label{noise-2} \\
 {\rm Re}\, \int\limits_{-\infty}^\infty  d\omega_{p}\, e^{i 2\pi \omega_{p} \frac{\delta \tau}{\tau} }\, S^{A*}_{out}(\epsilon,-\omega_{n}) S^{B}_{out}(\epsilon,-\omega_{m}) 
\nonumber \\
\times  S^{B*}_{out}(\epsilon + \omega_{p}, \omega_{p} - \omega_{m} )  S^{A}_{out}(\epsilon + \omega_{p}, \omega_{p} - \omega_{n})   \,, \nonumber
\end{eqnarray}
\ \\
\noindent
where ${\rm Re}$ indicates the real part of an expression and 

\begin{eqnarray}
\delta\tau = \tau_{L^{A}} - \tau_{L^{B}} \,.
\label{dtau}
\end{eqnarray}
Our aim here is to analyze how the shot noise depends on the difference of times when particles emitted by the different sources pass the QPC C. 
This difference depends on both the time when the particles were emitted and the time necessary for them to propagate to the QPC C. 
Without loss of generality we assume that the cavities A and B emit particles of the same kind (electrons or holes) at the same time. 
Therefore, in this subsection we use 

\begin{eqnarray}
S^{A}_{out} = S^{B}_{out} \equiv S_{out} \,.
\label{sym-col}
\end{eqnarray}
\ \\
\noindent
Thus $\delta\tau$ alone determines the difference of times when the particles pass the QPC C:
If $\delta\tau \gg \tau_{D}$ the particles pass the QPC C independently, whereas if $\delta\tau = 0$ they will collide.

\subsubsection{Independent particles}

If 

\begin{eqnarray}
\delta\tau \gg \tau_{D} \,,
\label{ind}
\end{eqnarray}
\ \\
\noindent
then we show that 

\begin{eqnarray}
{\cal P}_{34} = - 4 {\cal P}_{0} \,,
\label{ShNQE}
\end{eqnarray}
\ \\
\noindent
i.e., each particle contributes the same value $-{\cal P}_{0}$. 

To arrive at Eq.~(\ref{ShNQE}) we first represent the Fermi functions difference in Eq.~(\ref{noise-2}) as 

\begin{eqnarray}
\{f \left( \epsilon + \omega_{n}  \right)  - f \left( \epsilon + \omega_{m} \right) \}^2 = \{f \left( \epsilon + \omega_{n}  \right)  - f \left( \epsilon  \right) \}^2 \nonumber \\
+ \{f \left( \epsilon   \right)  - f \left( \epsilon + \omega_{m} \right) \}^2 \, \quad  \, \quad \,
\label{dfid} \\
+ 2 \{f \left( \epsilon + \omega_{n}  \right)  - f \left( \epsilon  \right) \} \{f \left( \epsilon   \right)  - f \left( \epsilon + \omega_{m} \right) \} \,. \nonumber
\end{eqnarray}
\ \\
\noindent
Then, for instance, with the first term on the right hand side  of Eq.~(\ref{dfid}) and with Eq.~(\ref{sym-col}) we can integrate out $\omega_{m}$ in Eq.~(\ref{noise-2}). 
Next we use the unitarity condition, \cite{Moskalets:2011cw}

\begin{eqnarray}
\int\limits_{-\infty}^{\infty} d\omega_{m}\, S_{out}^{*}(\epsilon + \omega_{p}, \omega_{p} - \omega_{m}) S_{out}(\epsilon + \omega_{q}, \omega_{q} - \omega_{m}) \nonumber \\
= \left( \frac{\hbar\Omega }{\Delta } \right)^{2} \delta(\omega_{p} - \omega_{q})\,. \, \quad \, \quad \, \label{uni-1-1} 
\end{eqnarray}
\ \\
\noindent
complementary to Eq.~(\ref{uni-1}) and get $\delta(\omega_{p})$.
After that we integrate out $\omega_{p}$, and arrive at an equation similar to Eq.~(\ref{noise-1}), which is shown to be equal to $-2{\cal P}_{0}$, see Eq.~(\ref{noise-1-2}). 
The same procedure with the second term on the right hand side of Eq.~(\ref{dfid}) results in a secondl contribution $-2{\cal P}_{0}$. 
To prove Eq.~(\ref{ShNQE}) we have to show additionally that what remains in Eq.~(\ref{noise-2}) is zero,  

\begin{eqnarray}
{\cal P}_{34 }^{rest}  = 2 {\cal P}_{0}\!\!    \int\limits_{-\infty}^\infty  d\epsilon\,  {\rm Re}\! \int\limits_{-\infty}^\infty  d\omega_{p}\, e^{i 2\pi \omega_{p} \frac{\delta\tau}{\tau} } \, |J(\epsilon,\omega_{p})|^{2}\,,  \nonumber \\
\label{noise-2-rest} 
\end{eqnarray}
\ \\
\noindent
where

\begin{eqnarray}
J(\epsilon,\omega_{p}) = \frac{\Delta^{2}}{(\hbar\Omega)^{2}} \int\limits_{-\infty}^{\infty} d\omega_{n}  \{f \left( \epsilon + \omega_{n}  \right)  - f \left( \epsilon  \right) \} \nonumber \\
\times  S^{*}_{out}(\epsilon, -\omega_{n})   S_{out}(\epsilon + \omega_{p}, \omega_{p} - \omega_{n}) \,. \nonumber
\end{eqnarray}
\ \\
\noindent
To show this we note that $|\delta\tau| \gg \tau$ as it follows from Eqs.~(\ref{ind}) and (\ref{taud}) for $T \ll 1$. 
We see that the integrand in Eq.~(\ref{noise-2-rest}) oscillates fast with $\omega_{p}$ and, therefore, the integral over $\omega_{p}$ is zero, ${\cal P}_{34 }^{rest} = 0$.

\subsubsection{Colliding particles}

If the particles collide at the QPC C,

\begin{eqnarray}
\delta\tau = 0 \,,
\label{col}
\end{eqnarray}
\ \\
\noindent 
then the noise is zero, 

\begin{eqnarray}
{\cal P}_{34} = 0 \,.
\label{ShNSE}
\end{eqnarray}
\ \\
\noindent 
This follows directly from Eq.~(\ref{noise-2}), where we use Eqs.~(\ref{sym-col}) and (\ref{uni-1}) and integrate over $\omega_{p}$. 
As a result we find $\delta(\omega_{n} - \omega_{m})$, which in turn vanishes after the integration, say, over $\omega_{n}$ due to the different Fermi functions. 

We emphasize that both Eqs.~(\ref{ShNQE}) and (\ref{ShNSE}) were obtained without any reference to the condition of emission.

\subsubsection{Partial overlap of wave packets}

Now we analyze how the noise depends on the overlap of particles at the QPC C. The overlap is a function of the time delay $\delta\tau \sim \tau_{D}$.
We use Eqs.~(\ref{sym-col}) and (\ref{out-n-1}) in Eq.~(\ref{noise-2}), which after the following substitutions, $\omega_{p} \to \omega_{p} - \epsilon$, $\omega_{n/m} \to - \omega_{n/m} - \epsilon$ can be cast into the following form:

\begin{eqnarray}
{\cal P}_{34 }^{na}  &=& -\, {\cal P}_{0}    \int\limits_{-\infty}^\infty  d\omega_{n} \int\limits_{-\infty}^\infty  d\omega_{m}\, |J(\omega_{n}, \omega_{m})|^{2}    \nonumber \\
&& \times \left\{f \left( - \omega_{n}  \right) - f \left( - \omega_{m} \right) \right\}^2 \,,  
\label{noise-2-na} 
\end{eqnarray}
\ \\
\noindent
where 

\begin{eqnarray}
J(\omega_{n}, \omega_{m}) = \int\limits_{-\infty}^\infty  d\omega_{p} e^{i 2\pi \omega_{p}  \frac{ \delta\tau }{\tau} } \, \quad \, \quad \,
\label{jnm}  \\
\times \frac{  \sin(\pi [\omega_{p} +  \omega_{n}])  }{\pi [\omega_{p} + \omega_{n} ]  }\,  \frac{  \sin(\pi [\omega_{p} + \omega_{m} ])  }{\pi [\omega_{p} + \omega_{m} ]  }\, \nonumber \\
\times \left\{ \delta(\omega_{p} + \omega_{n})   -   \frac{ \frac{ e^{i 2\pi [\omega_{p} + \omega_{n}] \frac{t_{-} }{\tau } }  }{\left(1 - \omega_{p} - \omega_{n} \right) } +  \frac{ e^{i 2\pi [\omega_{p} + \omega_{n}] \frac{t_{+} }{\tau } }  }{\left( 1 + \omega_{p} + \omega_{n}  \right) }  }{\rho^{*}(\omega_{p}) \rho(-  \omega_{n}) }  \right\} \nonumber \\
\times \left\{ \delta(\omega_{p} + \omega_{m} )   -   \frac{ \frac{e^{i 2\pi [\omega_{p} + \omega_{m}] \frac{t_{-} }{\tau } }  }{\left(1 -  \omega_{p} - \omega_{m}  \right) } +  \frac{e^{i 2\pi [\omega_{p} + \omega_{m}] \frac{t_{+} }{\tau } }  }{\left(1 + \omega_{p} + \omega_{m} \right) } }{\rho(\omega_{p}) \rho^{*}(-  \omega_{m}) }  \right\} . \nonumber 
\end{eqnarray}
\ \\
\noindent
To simplify the equation above we note the following. 
First, the term $\delta(\omega_{m} - \omega_{n})$ does not contribute to Eq.~(\ref{noise-2-na}). 
Second, since $|t_{-} - t_{+}| \sim {\cal T} \gg \tau$ then any term containing $\exp\{\omega_{p} [t_{-} - t_{+}]/\tau\}$ results in zero after the integration  over $\omega_{p}$. 
Furthermore after simple algebra we find,

\begin{eqnarray}
\!\!\!J(\omega_{n}, \omega_{m}) = \frac{{\cal A}_{-} e^{i 2\pi \omega_{q} \frac{t_{-}}{\tau} } + {\cal A}_{+} e^{i 2\pi \omega_{q}  \frac{t_{+}}{\tau} } }{ \rho^{*}(-\omega_{m}) \rho(-\omega_{n}) } \frac{ \sin (\pi \omega_{q} ) }{ \pi \omega_{q} } , \,
\label{jnm-1} 
\end{eqnarray}
\ \\
\noindent
where $\omega_{q} = \omega_{n} - \omega_{m}$ and 

\begin{eqnarray}
{\cal A}_{\mp} = \mp \frac{e^{-i 2\pi \omega_{n}\frac{\delta\tau }{\tau } } }{\omega_{q} \pm 1 } \pm \frac{e^{-i 2\pi \omega_{m}\frac{\delta\tau }{\tau } } }{\omega_{q} \mp 1 }  +  \hat {\cal A}_{\mp} \,,
\label{amp}
\end{eqnarray}
\ \\
\noindent
with

\begin{eqnarray}
\hat {\cal A}_{\mp} = \frac{  \pi \omega_{q}  }{\sin (\pi \omega_{q} )} \int\limits_{-\infty}^\infty  d\omega_{p}\frac{ e^{i 2\pi \omega_{p} \frac{\delta\tau }{\tau} } }{\pi^{2} \left | \rho(\omega_{p})  \right |^{2} } \, \quad \, \quad \,
\label{a-hat} \\
\times \frac{  \sin(\pi [\omega_{p} +  \omega_{n}])  \sin(\pi [\omega_{p} + \omega_{m} ])  }{ [\omega_{p} + \omega_{n} ] \left( \omega_{p} + \omega_{n}  \mp 1 \right) [\omega_{p} + \omega_{m} ] \left( \omega_{p} + \omega_{m}   \mp 1 \right)  } \,. \nonumber
\end{eqnarray}
\ \\
\noindent
We evaluate the last integral in the same way as we evaluated the integral in Eq.~(\ref{uni-1-4}): 

\begin{eqnarray}
\hat {\cal A}_{\mp} = \frac{  2  }{1 - \omega_{q}^{2} } \int\limits_{0}^{1}  d\omega_{p}^{\prime}\frac{ e^{i 2\pi \omega_{p}^{\prime} \frac{\delta\tau }{\tau} } }{ \left | \rho(\omega_{p}^{\prime})  \right |^{2} } \,.
\label{a-hat-1} 
\end{eqnarray}
\ \\
\noindent
Factorizing the original integral into the product of the integral over a single period and the sum over different periods we used the following prescription: The quantity $\hat A_{\mp}$ is a continuous function of $\delta\tau$ and it is changed only a little on the scale of $\tau$, which is the smallest time-scale in the problem and in many cases we put it to be zero, $\tau \to 0$. 
Therefore, we always consider $\delta\tau/\tau \gg 1$ to be, for instance, integer. 
As a consequence the integral in Eq.~(\ref{a-hat-1}) is the same for any period of the density of states $1/|\rho(\omega_{p})|^{2}$. 
Using Eq.~(\ref{BW}) we finally arrive at:

\begin{eqnarray}
\hat {\cal A}_{\mp} = \frac{  2 e^{- \frac{|\delta\tau| }{2\tau_{D} } }  }{1 - \omega_{q}^{2} }  \,.
\label{a-hat-2} 
\end{eqnarray}

To proceed we calculate $|J(\omega_{n},\omega_{m})|^{2}$ and keep only the terms with factors $|{\cal A}_{-}|^{2}$ and $|{\cal A}_{+}|^{2}$. 
All other terms, which have fast oscillating factors $e^{i 2\pi \omega_{q} t_{\mp}/\tau}$, will be nullified after the integration over $\omega_{n/m}$ in Eq.~(\ref{noise-2-na}). 
Thus,  

\begin{eqnarray}
{\cal P}_{34 }^{na}\!\!  =  -\, {\cal P}_{0}\!\!    \int\limits_{-\infty}^\infty\!\! \frac{  d\omega_{n} }{ |\rho( \omega_{n})|^{2} } \int\limits_{-\infty}^\infty \!\!\frac{  d\omega_{m} }{ |\rho( \omega_{m})|^{2} }  \left\{f \left( \omega_{n}  \right) - f \left( \omega_{m} \right) \right\}^2 \nonumber \\
\label{noise-2-na-1} \\
\times     \frac{4 \sin^{2} (\pi \omega_{q}) }{\pi^{2} \omega_{q}^{2} \left( 1 - \omega_{q}^{2} \right)^{2} }  \left\{ {\cal B}(\omega_{q})  - 4 e^{ - \frac{|\delta\tau| }{2\tau_{D} } }  \cos\left( 2\pi \omega_{n}\frac{\delta\tau }{\tau } \right)  \right\}  \,,
\nonumber 
\end{eqnarray}
\ \\
\noindent
where

\begin{eqnarray}
{\cal B}(\omega_{q}) = \omega_{q}^{2} + 1 + 2e^{- \frac{|\delta\tau| }{\tau_{D} }} - [\omega_{q}^{2} - 1 ]\cos\left( 2\pi \omega_{q} \frac{\delta\tau }{\tau } \right) . \nonumber \\
\label{bq}
\end{eqnarray}
\ \\
\noindent
Note in Eq.~(\ref{noise-2-na-1}) we changed the sign, $\omega_{n/m} \to -\omega_{n/m}$, compared to Eq.~(\ref{noise-2-na}). 
In addition we used the symmetry with respect to $\omega_{n}$ and $\omega_{m}$ and replaced  $\cos\left( 2\pi \omega_{n} \delta\tau/\tau  \right) + \cos\left( 2\pi \omega_{m} \delta\tau/ \tau  \right)$ by $2 \cos\left( 2\pi \omega_{n} \delta\tau/ \tau  \right)$.

Let us first evaluate the part of Eq.~(\ref{noise-2-na-1}) with ${\cal B}(\omega_{q})$.
We use Eq.~(\ref{BW}).
In the leading order in $g\to 0$ and at small temperatures, $f(\omega_{n}) \approx \theta(-\omega_{n})$, we find that only $\omega_{q} = \pm 1$ is relevant. 
Using ${\cal B}(\pm 1) = 2 \left[ 1 + \exp\left( - |\delta\tau|/\tau_{D} \right) \right]$ we calculate the corresponding contribution 

\begin{eqnarray}
{\cal P}_{34 }^{na, 1 } = - 4 {\cal P}_{0} \left\{ 1 + e^{- \frac{|\delta\tau|}{\tau_{D}}} \right\} \,.
\label{p1}
\end{eqnarray}
\ \\
\noindent
Similarly we evaluate the remaining part,

\begin{eqnarray}
{\cal P}_{34 }^{na,2} &=&  8 e^{ - \frac{|\delta\tau| }{2\tau_{D} } } {\cal P}_{0}  \int\limits_{-0.5}^{0.5}  d\omega_{n}\, \frac{g/\pi  \cos\left( 2\pi \omega_{n}\frac{\delta\tau }{\tau } \right) }{  \omega_{n} ^{2} + g^{2}}  \nonumber \\
&=& 8 e^{ - \frac{|\delta\tau| }{\tau_{D} } } {\cal P}_{0} \,.
\label{p2} 
\end{eqnarray}
\ \\
\noindent
Summing up Eqs.~(\ref{p1}) and (\ref{p2}) we arrive at

\begin{eqnarray}
{\cal P}_{34}^{na} =  - 4 {\cal P}_{0} \left\{ 1 - e^{- \frac{|\delta\tau|}{\tau_{D}}} \right\} \,.
\label{finnoise}
\end{eqnarray}
We see that at $|\delta\tau| \to \infty,$ we recover the independent particle case, see Eq.~(\ref{ShNQE}), whereas 
at $\delta\tau = 0$ the zero-noise result for colliding particles is recovered, see Eq.~(\ref{ShNSE}).
If electrons and holes have different time delays we arrive at Eq.~(\ref{sh-2}).

\section{Two-particle emitter}
\label{twope}

Let two cavities be coupled to the same edge state, a distance $L$ away each other (see Fig.~\ref{fig2}). 
We assume that the cavities emit particles at different times but are otherwise identical.

\subsection{The scattering amplitude}

The elements of the Floquet scattering matrix, $\hat S_{F}^{(2)}$, of the two-cavity system are expressed it terms of the elements of the Floquet scattering matrices of the cavities, $\hat S_{F}^{A}$ and $\hat S_{F}^{B}$, in the following way \cite{Moskalets:2011cw} 

\begin{eqnarray}
S_{F}^{(2)}(E,E_{n})\! =\!\! \sum\limits_{\ell=-\infty}^{\infty}\!\! S_{F}^{B}(E,E_{\ell})  e^{i \varphi_{L}(E)} e^{i \ell \Omega \tau_{L} } S_{F}^{A}(E_{\ell},E_{n}) , \nonumber \\
\label{tot}
\end{eqnarray}
\ \\
\noindent
where $\varphi_{L}(E)$ is the phase factor describing free propagation between the cavities, $\tau_{L}$ the time of flight from A to B.
Using Eqs.~(\ref{dimless}) and introducing $S_{out}^{j}$ we rewrite 

\begin{eqnarray}
S_{out}^{(2)}(\epsilon, - \omega_{n}) &=& \frac{\Delta}{\hbar\Omega} \int d\omega_{\ell}  \, e^{i\varphi_{L}(\epsilon)}\, e^{i 2\pi \omega_{\ell} \frac{\tau_{L}}{\tau}}
\label{tot-cfr} \\
&& \times S_{out}^{B}(\epsilon, - \omega_{\ell}) S_{out}^{A}(\epsilon + \omega_{\ell}, \omega_{\ell} - \omega_{n}) \,. \nonumber
\end{eqnarray}
\ \\
\noindent
We simplify the above equation in two important cases: (i) If the cavities emit an electron and a hole at close times and (ii) if they emits two electrons at close times. 
We use $S_{out}^{j}$,  Eq.~(\ref{out-n-1}) with $t_{\mp}$ replaced by $t_{\mp}^{j}$.
We will use the upper indices ``$eh$'' and ``$ee$'' to distinguish quantities related to these cases.

\subsubsection{Electron-hole emission}

We  keep the terms with $t_{-}^{A}$ and $t_{+}^{B}$ in equations for $S_{out}^{A}$ and $S_{out}^{B}$, respectively, and find

\begin{eqnarray}
S_{out}^{(2)eh}(\epsilon,-\omega_{n}) = e^{ i\left ( \psi(\epsilon)  + \pi \omega_{n} \right) } \frac{\hbar\Omega}{\Delta} \frac{\rho^{*}(\epsilon)  }{\rho(\epsilon) } \frac{\rho^{*}(\epsilon+\omega_{n})  }{\rho(\epsilon+\omega_{n}) } \nonumber \\
\label{tot-cfr-3} \\
\times  e^{- i 2\pi  \omega_{n} \frac{ t_{-}^{A} }{\tau } }  \left\{  \frac{ \delta S_{out}^{(2)eh}(\epsilon,-\omega_{n})  }{ \rho^{*}(\epsilon) \rho^{*}(\epsilon+\omega_{n})  } + \delta(\omega_{n})  +\frac{\sin(\pi \omega_{n}) }{\pi \omega_{n}}  \right. \nonumber \\
\left. \times   \left [  \frac{  e^{ i 2\pi \omega_{n}\frac{  \delta t^{eh} }{\tau }   } /\left( \omega_{n}  -1 \right)  }{ \rho^{*}(\epsilon) \rho(\epsilon + \omega_{n}) }   -  \frac{ 1/ \left( \omega_{n}  +1 \right)  }{  \rho^{*}(\epsilon+\omega_{n}) \rho(\epsilon ) }  \right] \right\}  \,, \nonumber
\end{eqnarray}
\ \\
\noindent
where $\psi(\epsilon) = 2\theta_{r} + 2\phi(\epsilon) + \varphi_{L}(\epsilon)$ and 

\begin{eqnarray}
\delta t^{eh} = t_{-}^{A} + \tau_{d} - t_{+}^{B} \,,
\label{dt}
\end{eqnarray}
and 

\begin{eqnarray}
\delta S_{out}^{(2)eh}(\epsilon,-\omega_{n}) = \int d\omega_{\ell}  \frac{  \sin(\pi  \omega_{\ell})  }{\pi \omega_{\ell}  } \frac{  \sin(\pi  [\omega_{\ell}-\omega_{n}])  }{\pi [\omega_{\ell} - \omega_{n}]  } \nonumber \\
\label{tot-cfr-4} \\
\times    \frac{  \exp\left\{ i 2\pi \omega_{\ell}\frac{ \delta t^{eh} }{\tau }   \right\}   }{\left( \omega_{\ell}  -1 \right) \left( \omega_{\ell}-\omega_{n}  -1 \right) \rho^{2}(\epsilon + \omega_{\ell}) }  \,. \nonumber 
\end{eqnarray}
\ \\
\noindent
Here in $\exp\left\{ i 2\pi \omega_{\ell} \delta t^{eh}/ \tau    \right\} $ we have neglected $\tau$ compared to $ \delta t^{eh}$.
To simplify further we expand $1/\rho^{2}$ into the sum of the Breit-Wigner resonances,

\begin{eqnarray}
\frac{1}{\rho^{2}(\epsilon+\omega_{\ell}) } =  \sum\limits_{a=-\infty}^{\infty} \frac{- g/\pi }{ (\epsilon + \omega_{\ell} + 0.5 - a + i g)^{2} } \,,
\label{rho-1}
\end{eqnarray}
\ \\
\noindent
and integrate out $\omega_{\ell}$ as we already did.
As a result we obtain:

\begin{eqnarray}
\delta S_{out}^{(2)eh}(\epsilon,-\omega_{n}) &=&  \theta(-\delta t^{eh}) \, e^{-\,\frac{|\delta t^{eh}| }{2\tau_{D} } }\, \frac{\delta t^{eh}  }{\tau_{D} }  \nonumber \\
&& \times   e^{ -i 2\pi  \epsilon  \frac{ \delta t^{eh} }{\tau }}\frac{2 \sin(\pi \omega_{n})  }{ \pi \omega_{n} \left( \omega_{n}^{2} - 1 \right)} \,. 
\label{tot-cfr-6} 
\end{eqnarray}
\ \\
\noindent
Interestingly, this term vanishes at $\delta t^{eh} = 0$, when an electron and a hole are emitted simultaneously, as well as at $|\delta t^{eh}| \gg \tau_{D}$, when they are emitted independently.

\subsubsection{Two electron emission}

Keeping the terms with $t_{-}^{A}$ and $t_{-}^{B}$ in equations for $S_{out}^{A}$ and $S_{out}^{B}$, respectively, we calculate  

\begin{eqnarray}
S_{out}^{(2)ee}(\epsilon, -\omega_{n}) = e^{ i\left ( \psi(\epsilon)  + \pi \omega_{n} \right) } \frac{\hbar\Omega}{\Delta} \frac{\rho^{*}(\epsilon)  }{\rho(\epsilon) } \frac{\rho^{*}(\epsilon+\omega_{n})  }{\rho(\epsilon+\omega_{n}) } \nonumber \\
\times  e^{- i 2\pi  \omega_{n} \frac{ t_{-}^{A} }{\tau } }  \left\{  \frac{ \delta S_{out}^{(2)ee}(\epsilon, -\omega_{n})  }{ \rho^{*}(\epsilon) \rho^{*}(\epsilon+\omega_{n})  } + \delta(\omega_{n})    \right. \, \quad \, \quad \,
\label{tot-cfr-3ee} \\
\left. -\frac{\sin(\pi \omega_{n}) }{\pi \omega_{n} \left( \omega_{n}  + 1 \right) }   \left [  \frac{  e^{ i 2\pi \omega_{n}\frac{  \delta t^{ee} }{\tau }   }   }{ \rho^{*}(\epsilon) \rho(\epsilon + \omega_{n}) }  +  \frac{ 1   }{ \rho^{*}(\epsilon+\omega_{n}) \rho(\epsilon ) }  \right] \right\}  \,, \nonumber
\end{eqnarray}

\begin{eqnarray}
\delta S_{out}^{(2)ee}(\epsilon, -\omega_{n}) &=&  \theta(-\delta t^{ee})  \,e^{-\,\frac{|\delta t^{ee}| }{2\tau_{d} } }\, \frac{\delta t^{ee}  }{\tau_{D} }  \nonumber \\
&&\times e^{ -i 2\pi  \epsilon  \frac{ \delta t^{ee} }{\tau }} \frac{2 \sin(\pi \omega_{n}^{\prime})  }{ \pi \omega_{n}^{\prime} \left( 1 - \omega_{n}^{\prime 2} \right)} \,, 
\label{tot-cfr-6ee} 
\end{eqnarray}
\ \\
\noindent
where $\omega_{n}^{\prime} = \omega_{n} + 1$ and

\begin{eqnarray}
\delta t^{ee} = t_{-}^{A} + \tau_{d} - t_{-}^{B} \,.
\label{dtee}
\end{eqnarray}
\ \\
\noindent
The term $\delta S_{out}^{(2)ee}(\epsilon, -\omega_{n})$ is irrelevant in both the case of emission of independent electrons, $|\delta t^{ee}| \gg \tau_{D}$, and in the case of emission of a pair of electrons, $\delta t^{ee} = 0$.

\subsection{DC heat flow}

The excess energy, i.e., the energy over the Fermi energy, carried out by the particles emitted during each period can be calculated as a DC heat flow, $I_{DC}^{Q}$, which is expressed in terms of the Floquet scattering amplitude as follows \cite{MBstrong02,Moskalets:2011cw}

\begin{eqnarray}
I_{DC}^{Q}   & = &  \dfrac{\Delta^{3}}{h\hbar\Omega} \int\limits_{-\infty}^{\infty} d\omega_{n} \int\limits_{-\infty}^{\infty} \epsilon\, d\epsilon  \left\{ f  \left( \epsilon + \omega_{n} \right) - f \left( \epsilon  \right) \right\}  \nonumber \\
&&\times  \left |  S_{out}^{(2)}\left(\epsilon, -\omega_{n}  \right)  \right |^{2}\,. 
\label{nah01-2} 
\end{eqnarray}
\ \\
\noindent
We use this equation to analyze different conditions of emission.

\subsubsection{Emission of independent particles} 
\label{heatind}

We use Eq.~(\ref{tot-cfr-3}) at $\delta t^{eh} \gg \tau_{D}$ in Eq.~(\ref{nah01-2}) and obtain

\begin{eqnarray}
I_{DC}^{Q}   =   \dfrac{\Delta}{\cal T} \int\limits_{-\infty}^{\infty} d\omega_{n} \int\limits_{-\infty}^{\infty} \epsilon\,  d\epsilon\, \frac{ f  \left( \epsilon + \omega_{n}  \right) - f \left( \epsilon   \right) }{|\rho(\epsilon )|^{2} | \rho(\epsilon + \omega_{n} )|^{2} }  \nonumber \\
\label{nah01-4} \\
\times  \frac{\sin^{2}(\pi \omega_{n}) }{\pi^{2} \omega_{n}^{2} }  \left\{  \frac{  1 }{\left( \omega_{n}  -1 \right)^{2} } +  \frac{ 1   }{\left( \omega_{n}  +1 \right)^{2}  }  \right\}  \,. \nonumber 
\end{eqnarray}
\ \\
\noindent
Here we dropped a term proportional to $\exp\left( i 2\pi \omega_{n} \delta t^{eh}/\tau \right)$ since it is oscillating fast and hence vanishes after the integration over $\omega_{n}$.
To continue we use Eq.~(\ref{BW}) to integrate over $\omega_{n}$ and over $\epsilon$ and find: 

\begin{eqnarray}
I_{DC}^{Q}   =   \dfrac{\Delta}{\cal T} \,.
\label{nah01-5}
\end{eqnarray}
\ \\
\noindent
Note  that both terms in the curly brackets in Eq.~(\ref{nah01-4}) contribute equally. 
The first one corresponds to a hole emission and the second one corresponds to an electron emission. 
Therefore, each particle carries an energy ${\cal E}_{na} = \Delta/2$, see Eq.~(\ref{en-na}). 
The same answer, Eq.~(\ref{nah01-5}), is obtained if we use Eq.~(\ref{tot-cfr-3ee}) at $\delta t^{ee} \gg \tau_{D}$.

Notice, since both cavities together emit four particles per period, two electrons and two holes, the total DC heat current is twice as large as  $I_{DC}^{Q}$, Eq.~(\ref{nah01-5}). 

\subsubsection{Electron-hole pair emission}
\label{heateh}

If $\delta t^{eh} = 0$, then Eq.~(\ref{nah01-2}) with $S_{out}^{(2)eh}$ from Eq.~(\ref{tot-cfr-3}) gives

\begin{eqnarray}
I_{DC}^{Q}   =   \dfrac{\Delta}{\cal T} \int\limits_{-\infty}^{\infty} d\omega_{n} \int\limits_{-\infty}^{\infty} \epsilon\, d\epsilon \frac{ f  \left( \epsilon + \omega_{n} \right) - f \left( \epsilon  \right) }{|\rho(\epsilon)|^{2} | \rho(\epsilon+\omega_{n})|^{2} }  \nonumber \\
\label{nah01-6} \\
\times \frac{\sin^{2}(\pi \omega_{n}) }{\pi^{2} \omega_{n}^{2} }  \left\{  \frac{  1 }{\left( \omega_{n}  -1 \right)^{2} } +  \frac{ 1   }{\left( \omega_{n}  +1 \right)^{2}  } - \frac{2\xi(\epsilon,\omega_{n}) }{\omega_{n}^{2} - 1 }  \right\}  \,, \nonumber 
\end{eqnarray}
\ \\
\noindent
where

\begin{eqnarray}
\xi(\epsilon,\omega_{n}) = {\rm Re} \frac{\rho^{*}(\epsilon + \omega_{n}) \rho(\epsilon) }{\rho(\epsilon+\omega_{n}) \rho^{*}(\epsilon) }\,.
\label{chi}
\end{eqnarray}
\ \\
\noindent
Evaluating integrals over $\epsilon$ and $\omega_{n}$ at $g\to 0$ along the same lines as before we find that the term proportional to $\xi(\epsilon,\omega_{n})$ does not contribute. 
Hence, Eq.~(\ref{nah01-5}) remains valid even if an electron and a hole are emitted simultaneously. 
Therefore, there is no re-absorption\cite{SOMB08,MB09} under the non-adiabatic emission condition.

\subsubsection{Electron-electron pair emission} 
\label{heatee}

We substitute Eq.~(\ref{tot-cfr-3ee}) into Eq.~(\ref{nah01-2}) and find at $\delta t^{ee} = 0$:

\begin{eqnarray}
I_{DC}^{Q,ee}   &=&   \dfrac{\Delta}{\cal T} \int\limits_{-\infty}^{\infty} d\omega_{n} \int\limits_{-\infty}^{\infty} \epsilon\, d\epsilon \frac{ f  \left( \epsilon + \omega_{n} \right) - f \left( \epsilon  \right) }{|\rho(\epsilon)|^{2} | \rho(\epsilon+\omega_{n})|^{2} }  \nonumber \\
&& \times \frac{2 \sin^{2}(\pi \omega_{n}) [1 + \xi(\epsilon,\omega_{n}) ] }{\pi^{2} \omega_{n}^{2} \left( \omega_{n}  +1 \right)^{2} }  \,. \label{nah01-6ee}  
\end{eqnarray}
\ \\
\noindent
With $1/|\rho(\epsilon)|^{2}$ given by Eq.~(\ref{BW}) we calculate at $g\to 0$:

\begin{eqnarray}
I_{DC}^{Q,ee}   =   \dfrac{2\Delta}{\cal T} \,.
\label{nah01-5ee}
\end{eqnarray}
\ \\
\noindent
We see that the energy carried by the two-electron pair, ${\cal E}_{na}^{ee} = 2\Delta$, is as twice as large as the energy, $2{\cal E}_{na} = \Delta$, carried by two electrons emitted separately.
This energy enhancement is due to the Pauli exclusion principle which forbids that the two electrons emitted simultaneously have the same energy.

The doubling of the work done by the two-particle source when it emits two-electron pairs is a peculiar feature common to both adiabatic \cite{MB09} and non-adiabatic, Eq.~(\ref{nah01-5ee}), conditions of emission.


\begin{thebibliography}{11}


\bibitem{Moskalets:2011cw}
M. V. Moskalets, {\it Scattering matrix approach to non-stationary quantum transport} (ICP, London, 2011). 

\bibitem{Feve:2007jx}
G. F\`{e}ve, A. Mah\'e, J.-M. Berroir, T. Kontos, B. Pla\c{c}ais, D. C. Glattli, A. Cavanna, B. Etienne, Y. Jin, 
Science {\bf 316}, 1169 (2007).

\bibitem{Blumenthal07}
M. D. Blumenthal, B. Kaestner, L. Li, S. Giblin, T. J. B. M. Janssen, M. Pepper, D. Anderson, G. Jones, and D. A. Ritchie,
Nature Physics {\bf 3}, 343 (2007).

\bibitem{Maire:2008hx}
N. Maire, F. Hohls, B. Kaestner, K. Pierz, H. W. Schumacher, and R. J. Haug, Appl. Phys. Lett. {\bf 92}, 082112 (2008).

\bibitem{Mahe:2010cp}
A. Mah\'{e},  F. D. Parmentier,  E. Bocquillon,  J.-M. Berroir,  D. C. Glattli, T. Kontos,  B. Pla\c{c}ais, G. F\`{e}ve,  A. Cavanna,  and Y. Jin, 
Phys. Rev. B {\bf 82}, 201309 (2010).

\bibitem{Bocquillon:2012if}
E. Bocquillon, F. D. Parmentier, C. Grenier, J.-M. Berroir, P. Degiovanni, D. C. Glattli, B. Pla\c{c}ais, A. Cavanna, Y. Jin, and G. F\`eve, 
Phys. Rev. Lett. {\bf 108}, 196803 (2012).

\bibitem{Fricke:2012vk}
L. Fricke, M. Wulf, B. Kaestner, V. Kashcheyevs, J. Timoshenko, P. Nazarov, F. Hohls, P. Mirovsky, B. Mackrodt, R. Dolata, T. Weimann, K. Pierz, and H. W. Schumacher, arXiv:1211.1781 (upublished).

\bibitem{Grenier11}
Ch. Grenier, R Herv\'{e}, E. Bocquillon, F. D. Parmentier, B Pla\c{c}ais, J. M. Berroir, G F\`{e}ve,  and P. Degiovanni, New J. Phys. {\bf 13}, 093007 (2011).

\bibitem{Grenier:2011js}
C. Grenier, R. and Herv\'{e}, G. F\`{e}ve, P. Degiovanni, Int. J. Mod. Phys. B {\bf 25}, 1053 (2011).

\bibitem{HMB12}
G. Haack, M. Moskalets, and M. B\"uttiker, 
arXiv:1212.0088 (unpublished).

\bibitem{Fujiwara08}
A. Fujiwara, K. Nishiguchi, Y. Ono,
Appl. Phys. Lett. {\bf 92}, 042102 (2008).

\bibitem{Kaestner08}
B. Kaestner, V. Kashcheyevs, G. Hein, K. Pierz,  U. Siegner, and H. W. Schumacher,
Appl. Phys. Lett. {\bf 92}, 192106 (2008).

\bibitem{Wright11}
S. J. Wright,  A. L. Thorn, M. D. Blumenthal,  S. P. Giblin,  M. Pepper,  T. J. B. M. Janssen,  M. Kataoka,  J. D. Fletcher,  G. A. C. Jones,  C. A. Nicoll, G. Gumbs,  and D. A. Ritchie,
J. Appl. Phys. {\bf 109}, 102422 (2011). 

\bibitem{Leich11}
C. Leicht, P. Mirovsky , B. Kaestner, F. Hohls, V. Kashcheyevs, E. V. Kurganova, U. Zeitler, T. Weimann, K. Pierz  and H. W. Schumacher,
Semicond. Sci. Technol. {\bf 26}, 055010  (2011). 

\bibitem{Fletcher:2012te}
J. D. Fletcher, M. Kataoka, H. Howe, M. Pepper, P. See, S. P. Giblin, J. P. Griffiths, G. A. C. Jones, I. Farrer, D. A. Ritchie, and T. J. B. M. Janssen, 
arXiv:1212.4981v1 (unpublished).

\bibitem{BS11}
F. Battista and P. Samuelsson,
Phys. Rev. B {\bf 83}, 125324 (2011).

\bibitem{Kashcheyevs:2012km}
V. Kashcheyevs and J. Timoshenko, Phys. Rev. Lett. {\bf 109}, 216801 (2012).

\bibitem{OSMB08} 
S. Ol'khovskaya, J. Splettstoesser, M. Moskalets, and M. B\"uttiker, 
Phys. Rev. Lett {\bf 101}, 166802 (2008).

\bibitem{KKL06}
J. Keeling, I. Klich, and L. S. Levitov,
Phys. Rev. Lett. {\bf 97}, 116403 (2006).

\bibitem{KSL08}
J. Keeling, A. V. Shytov, and L. S. Levitov, 
Phys. Rev. Lett. {\bf 101}, 196404 (2008).

\bibitem{ZSAM09}
J. Zhang, Y. Sherkunov, N. d'Ambrumenil, and B. Muzykantskii,
Phys. Rev. B {\bf 80}, 245308 (2009).

\bibitem{MB11}  
M. Moskalets and M. B\"{u}ttiker,  
Phys. Rev. B {\bf 83}, 035316 (2011). 

\bibitem{SMB09}
J. Splettstoesser, M. Moskalets, and M. B\"uttiker, 
Phys. Rev. Lett. {\bf 103}, 076804 (2009).

\bibitem{SSMB10}
J. Splettstoesser, P. Samuelsson, M. Moskalets, and M. B\"{u}ttiker,
J. Phys. A: Math. Theor. {\bf 43}, 354027 (2010).

\bibitem{SASB11}
Y. Sherkunov, N. d'Ambrumenil, P. Samuelsson, and M. B\"{u}ttiker, Phys. Rev. B {\bf 85}, 081108 (2012).

\bibitem{Sherkunov:2012dg}
C. H. Bennett and D. P. DiVincenzo, 
Nature (London) {\bf 404}, 247 (2000).

\bibitem{SOMB08} 
J. Splettstoesser, S. Ol'khovskaya, M. Moskalets, and M. B\"uttiker, 
Phys. Rev. B {\bf 78}, 205110 (2008).

\bibitem{MB09} 
M. Moskalets and M. B\"{u}ttiker,  
Phys. Rev. B {\bf 80}, 081302(R) (2009). 

\bibitem{JSMoskalets:2011cw}
S. Juergens, J. Splettstoesser, and M. Moskalets,
Europhys. Lett. {\bf 96}, 37011 (2011).

\bibitem{Levitov:1996ie}
L. S. Levitov, H. Lee, and G. B. Lesovik, J. Math. Phys. {\bf 37}, 4845 (1996).

\bibitem{Ivanov:1997wz}
D. A. Ivanov, H. W.  Lee, and L.S. Levitov, Physical Review B {\bf 56}, 6839 (1997).

\bibitem{Dubois:2012us}
J. Dubois, T. Jullien, C. Grenier, P. Degiovanni, P. Roulleau, and D. C. Glattli, 
arXiv:1212.3921 (unpublished).

\bibitem{Dubois12}
J. Dubois, T. Jullien, P. Roulleau, F. Portier, P. Roche, W. Wegscheider, and D. C. Glattli, (in preparation).

\bibitem{Gabelli06}
J. Gabelli, G. F\`{e}ve, J.-M. Berroir, B. Pla\c{c}ais,  A. Cavanna, B. Etienne, Y. Jin, D. C. Glattli,
Science {\bf 313}, 499 (2006).

\bibitem{Parmentier11}
F. D. Parmentier, E. Bocquillon, J.-M. Berroir, D. C. Glattli, B. Pla\c{c}ais and G. F\`{e}ve, and M. Albert, C. Flindt and M. B\"{u}ttiker,
Physical Review B {\bf 85}, 165438 (2012).


\bibitem{PTB96}         
A. Pr\^{e}tre, H. Thomas and M. B\"{u}ttiker, 
Phys. Rev. B {\bf 54}, 8130 (1996).

\bibitem{MSB08}
M. Moskalets, P. Samuelsson, and M. B\"{u}ttiker, 
Phys. Rev. Lett. {\bf 100}, 086601 (2008).    

\bibitem{BTP93}
M. B\"{u}ttiker, H. Thomas, and A. Pr\^{e}tre, 
Phys. Lett. A {\bf 180}, 364 (1993).

\bibitem{Sasaoka10}
K. Sasaoka, T. Yamamoto, and S. Watanabe,
Appl. Phys. Lett. {\bf 96}, 102105 (2010).

\bibitem{HMSB11}
G. Haack, M. Moskalets, J. Splettstoesser,  and M. B\"uttiker,
Phys. Rev. B {\bf 84}, 081303(R) (2011). 

\bibitem{Haack_these}
G. Haack, Th\`ese de doctorat, Universit\'e de Gen\`eve (2012).

\bibitem{Buttiker:1992vr}
M. B\"uttiker, Phys. Rev. B {\bf 46}, 12485 (1992).

\bibitem{Liu:1998wr}
R. Liu, B. Odom, Y. Yamamoto, and S. Tarucha, Nature {\bf 391}, 263 (1998).

\bibitem{Blanter:2000wi}
Y. M. Blanter and M. B\"{u}ttiker, Phys. Rep. {\bf 336}, 1 (2000).

\bibitem{Feve:2008im}
G. F\`eve, P. Degiovanni, and T. Jolicoeur, Physical Review B {\bf 77}, 035308 (2008).

\bibitem{MBnoise04} 
M. Moskalets and M. B\"{u}ttiker,  
Phys. Rev. B {\bf 70}, 245305 (2004). 

\bibitem{Hong:1987gm}
C. K. Hong, Z. Y. Ou, and L. Mandel, Phys. Rev. Lett. {\bf 59}, 2044 (1987).

\bibitem{Jonckheere:2012cu}
T. Jonckheere, J. Rech, C. Wahl, and T. Martin, Phys. Rev. B {\bf 86}, 125425 (2012).

\bibitem{Loudon}
R. Loudon, in {\it Disorder in Condensed Matter Physics}, eds. J. A. Blackman and J. Taguena (Clarendon Press, Oxford, 1991), p. 441.


\bibitem{Bocquillon:2013dp}
E. Bocquillon, V. Freulon, J.-M. Berroir, P. Degiovanni, B. Pla\c{c}ais, A. Cavanna, Y. Jin, and G. F\`{e}ve, Science, 24 January 2013 (10.1126/science.1232572).

\bibitem{Nigg:2006kl}
S. E. Nigg, R. Lopez, and M. B\"uttiker, 
Phys. Rev. Lett. {\bf 97}, 206804 (2006).

\bibitem{Petitjean:2009je}
C. Petitjean, D. Waltner, J. Kuipers, I. Adagideli, and K. Richter, Physical Review B {\bf 80}, 115310 (2009).

\bibitem{Mora:2010hw}
C. Mora and K. Le Hur, Nature Physics {\bf 6}, 697 (2010).

\bibitem{Kashuba:2012fs}
O. Kashuba, H. Schoeller, and J. Splettstoesser, Europhys. Lett. {\bf 98}, 57003 (2012).


\bibitem{Nigg08}
S. E. Nigg  and M. B\"{u}ttiker, 
Phys. Rev. B {\bf 77}, 085312 (2008). 

\bibitem{Avron:2001fl}
J. E. Avron, A. Elgart, G. M. Graf, and L. Sadun, Phys. Rev. Lett. {\bf 87}, 236601 (2001);
J. Math. Phys. {\bf 43}, 3415 (2002).


\bibitem{Kataoka11}
M. Kataoka, J. D. Fletcher, P. See, S. P. Giblin, T. J. B. M. Janssen,  J. P. Griffiths, G. A. C. Jones,  I. Farrer,  and D. A. Ritchie,
Phys. Rev. Lett. {\bf 106}, 126801 (2011).    

\bibitem{Battista:2012db}
F. Battista and P. Samuelsson, Phys. Rev. B {\bf 85}, 075428 (2012).

\bibitem{Arrachea:2007ih}
L. Arrachea, M. Moskalets, and L. Martin-Moreno, 
Phys. Rev. B {\bf 75}, 245420 (2007).

\bibitem{Arrachea:2012vb}
L. Arrachea and B. Rizzo, 
Journal of Physics: Conference Series, Progress in Nonequilibrium Green's Functions V (2012); arXive:1212.3181.

\bibitem{MB08} 
M. Moskalets and M. B\"{u}ttiker,  
Phys. Rev. B {\bf 78}, 035301 (2008). 

\bibitem{Buttiker:1994vl}
M. B\"uttiker, H. Thomas, and A. Pr\^{e}re, 
Z. Phys. B  {\bf 94}, 133 (1994).

\bibitem{Brouwer:1998jt}
P. W. Brouwer, 
Phys. Rev. B {\bf 58}, 10135 (1998).

\bibitem{MBstrong02} 
M. Moskalets and M. B\"{u}ttiker,  
Phys. Rev. B {\bf 66}, 205320 (2002). 

    


\end{thebibliography}
\end{document}